\documentclass[num-refs]{wiley-article}

\usepackage{siunitx}

\usepackage{makeidx} 
\usepackage{graphicx}
\usepackage{amsmath}
\usepackage{listings}
\usepackage{array}
\usepackage[colorinlistoftodos]{todonotes}
\usepackage{adjustbox}
\usepackage{svg}

\usepackage{booktabs}
\usepackage{geometry}

\papertype{Original Article}
\paperfield{Journal Section}

\title{Profiling and optimization of multi-card GPU machine learning jobs}

\abbrevs{HPC, High-Perfornce Computing; LLM, large language model; ML, machine learning.}

\author[1]{Marcin Lawenda}
\author[1]{Kyrylo Khloponin}
\author[1]{Krzesimir Samborski}
\author[2]{Łukasz Szustak}

\affil[1]{Data Processing Technologies Division, Poznan Supercomputing and Networking Center, Poznań, 61-139, Poland}
\affil[2]{Faculty of Mechanical Engineering and Computer Science, Czestochowa University of Technology, Częstochowa, 42-201, Poland}

\corraddress{Marcin Lawenda, Poznan Supercomputing and Networking Center, Jana Pawła II 10, Poznań, 61-139, Poland}
\corremail{lawenda@man.poznan.pl}

\fundinginfo{Funded by the European Union, European High Performance Computing Joint Undertaking and Poland, Germany, Spain, Hungary, France and Greece under grant agreement number: 101093457.}

\runningauthor{Lawenda et al.}

\begin{document}

\begin{frontmatter}
\maketitle

\begin{abstract}
The effectiveness and efficiency of machine learning methodologies are crucial, especially with respect to the quality of results and computational cost. This paper discusses different model optimization techniques, providing a comprehensive analysis of key performance indicators. Several parallelization strategies for image recognition, adapted to different hardware and software configurations, including distributed data parallelism and distributed hardware processing, are analyzed. Selected optimization strategies are studied in detail, highlighting the related challenges and advantages of their implementation. Furthermore, the impact of different performance improvement techniques (DPO, LoRA, QLoRA, and QAT) on the tuning process of large language models is investigated. Experimental results illustrate how the nature of the task affects the iteration time in a multiprocessor environment, VRAM utilization, and overall memory transfers. Test scenarios are evaluated on the modern NVIDIA H100 GPU architecture.

% Please include a maximum of seven keywords
\end{abstract}
\end{frontmatter}

%%%%%%%%%%%%%%%%%%%%%%%%%%%%%%%%%%%%%%%%%%%%%%%%%%%%%%%%%%%%
\section{Introduction}
Artificial intelligence methods, especially machine learning, are the subject of many discussions and studies. Their authors focus mainly on developing new models for analysing phenomena, spending much less time on optimization techniques, especially general-purpose ones that can be used in many implementations. The importance of the issue of efficient use of available resources is evidenced by the fact that the estimated energy consumption for data centres will reach 21\% in 2030 (from 2\% currently) of the total global energy demand, constituting a significant burden for the electric grid \cite{MITSloan}. 

This is particularly applicable in domains like machine learning (ML) and large language models (LLM), which have significantly impacted advancements across various fields, including natural language processing, computer vision, and data analysis \cite{10.1007/978-3-031-70445-1_29, 10710677, 8259424}. These models are characterized by their extensive parametrization and intricate architectures, necessitating substantial computational resources for both training and inference. Multi-card GPU systems, which utilize the parallel processing capabilities of several graphics processing units (GPUs), have become crucial infrastructure to fulfil these computational demands \cite{lin2024universalperformancemodelingmachine, 10.1145/3688351.3689164}. This strategy has led to the creation of code tailored for parallelization, facilitating the accelerated training of significantly larger models by harnessing the enhanced computational power and memory provided by additional accelerators. Nevertheless, the efficient allocation of tasks within multi-card GPU systems continues to pose a considerable challenge, owing to the complex nature of model architectures, data dependencies, and the variability of hardware \cite{10.1145/1058129.1058148}.

A major challenge for many AI applications is improving their performance, either through scalability or algorithmic improvements \cite{6339605}. One effective approach to addressing these challenges and implementing complex tasks is to run machine learning or LLM algorithms on specialized hardware accelerators. Adapting an application to a graphics processing unit (GPU) that was originally designed for the CPU can provide significant performance improvements, but this requires in-depth knowledge of the code itself and the hardware capabilities \cite{5289193,8782524}. This involves using various techniques related to process allocation and data management based on lessons learned from previous experiments. 

Determining the most appropriate method for a particular application workflow while maintaining the integrity of the results presents a significant challenge, particularly in environments utilizing multiple GPU cards across nodes. Consequently, the main aim of this study is to examine various optimization strategies, including data fragmentation, computational precision, memory management, and parameter tuning, in both single and multi-GPU settings. This analysis seeks to improve the processing efficiency of machine learning applications within the realm of AI development. The experiments conducted will assess the scalability and practicality of different techniques, ultimately facilitating enhanced planning and providing recommendations for related applications and advancements.

%%%%%%%%%%%%%%%%%%%%%%%%%%%%%%%%%%%%%%%%%%%%%%%%%%%%%%%%%%%%
\section{Related works}
The implementation of effective AI simulations requires a comprehensive understanding of the mechanisms by which they rely. Therefore, different techniques are used to evaluate performance and look for improvements, depending on the architecture of the test environment and the models used.

The article \cite{Pallipuram2012} presents an extensive analysis that integrates programming models, architectures, and applications. It compares four precise Spiking Neural Network models (Hodgkin-Huxley, Machine Learning, Wilson, and Izhikevich) within a two-tier character recognition network, distinguished by varying computation to communication ratios. Performance evaluations were conducted on Nvidia Fermi and AMD/ATi Radeon architectures utilizing the OpenCL (Open Computing Language) and CUDA programming models. Subsequently, optimization strategies were implemented, and the problem size was adjusted. The findings indicated that CUDA exhibited superior performance on the Nvidia GPU architecture in comparison to OpenCL. In most instances, both kernel invocations and memory transfers were more efficient with CUDA. The performance disparity is primarily attributed to the differing compilers employed by the two programming models. Nonetheless, there are instances where OpenCL may surpass CUDA, particularly in specific implementations of the Hodgkin-Huxley model. Notably, significant application speedups of up to 1095 times for the most computationally demanding SNN neuron model have been documented when compared to a serial implementation on an Intel Core 2 Quad host.

%%%%Model optimization

The paper \cite{appleyard2016optimizingperformancerecurrentneural} focuses on an optimization that achieves an order of magnitude speedup in network size by exploiting the parallelism between network operations compared to a naive implementation. The optimization is divided into three steps (first, single-cell optimization, second, single-layer optimization, and third, whole network optimization), which are part of the cuDNN (CUDA Deep Neural Network) library provided by NVIDIA. The effect of various factors on the inference time of a Long Short-Term Memory (LSTM) network was examined, specifically focusing on hidden state size, minibatch size, and the number of layers within the network. The results highlight the influence of optimization techniques on the forward pass of a four-layer LSTM network, which operates over 1000 steps, utilizes a hidden state size of 512, and processes an input minibatch of 64. In this scenario, an approximate speedup of 11 times is realized in comparison to a basic implementation, and a speedup of around 6 times is noted when contrasted with an implementation that employs standard GEMM (GEneral Matrix-matrix Multiplication) clustering optimization.

%%%%Parallel data distribution
The work \cite{10.14778/3415478.3415530} outlines the design, implementation and assessment of the PyTorch distributed data processing module. It discusses the effects of employing native PyTorch methods to enhance distributed data processing, such as gradient clustering, the overlap of computation with communication, and the omission of gradient synchronization. DDP (DistributedDataParallel), which facilitates parallel training across multiple processes and machines, accelerates this process by aggregating gradients into communication containers, overlaying communication with computations, and skipping synchronization. The findings indicate that, when configured appropriately, the PyTorch distributed data processing module can attain nearly linear scalability with the utilization of 256 GPUs. Furthermore, the experiments conducted indicated that the backpropagation step in DDP represents the most resource-intensive phase of training. This necessitates collaboration from both framework developers, who must facilitate the optimization algorithms, and application developers, who are responsible for empirically adjusting the parameters.

%%%%%%%%%%%%%%%%%%%%%%%%%%%%%%%%%%%%%%%%%%%%%%%%%%%%%%%%%%%%
\section{Experimental environment and methodology}\label{section:Methodology}

\subsection{Hardware infrastructure}
\begin{figure}[h]
    \centering
    \includegraphics[width=0.5\linewidth]{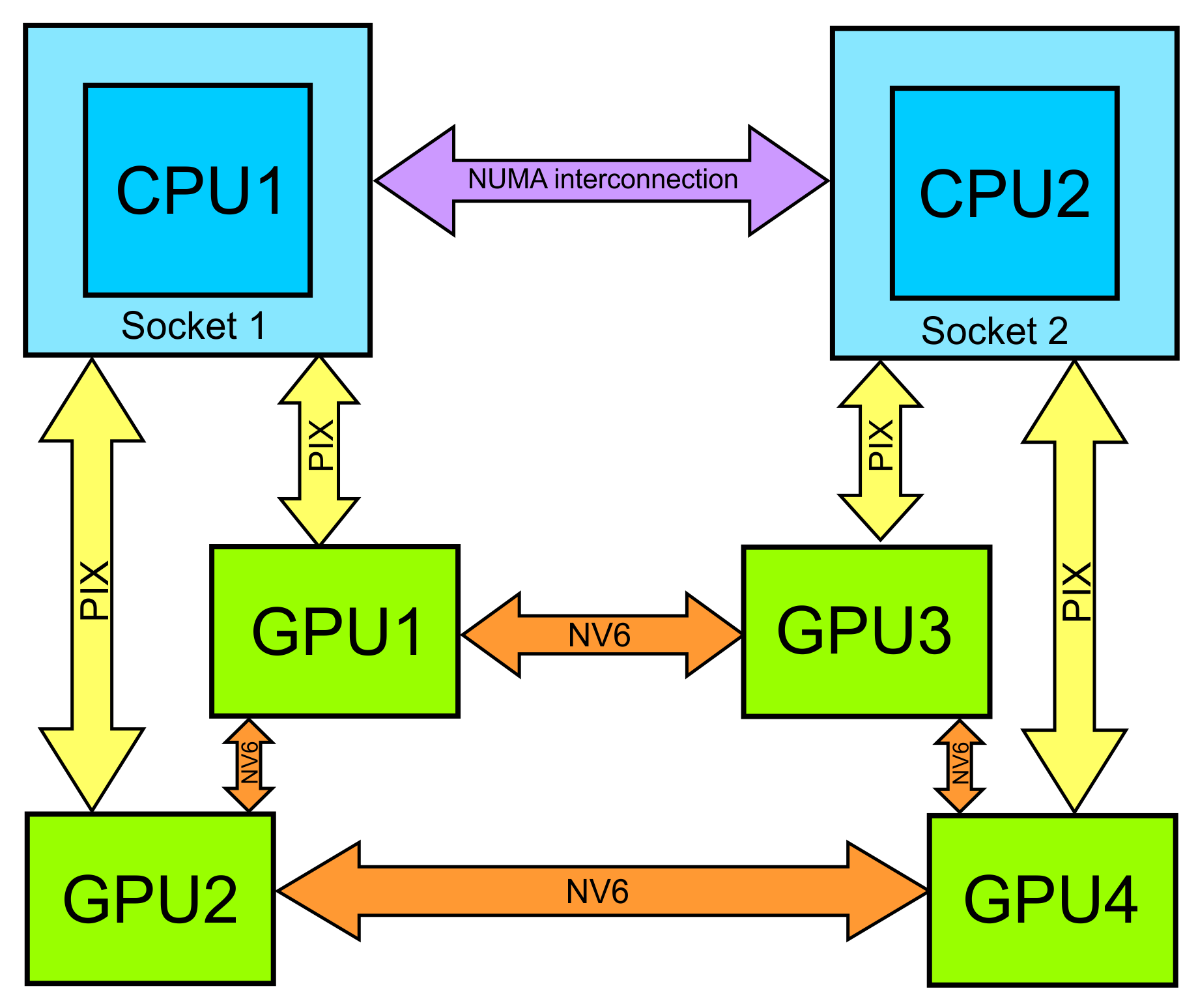}
    \caption{NUMA architecture of PROXIMA cluster nodes}
    \label{fig:numa_arch}
\end{figure}

The hardware platform utilized for this research is part of the HPC system called Proxima hosted at Poznan Supercomputing and Networking Center \cite{PSNC}. It is built on the HPE Cray XD665 system, which comprises 2 AMD EPYC 9334 CPUs and 4 NVIDIA H100-94 SXM5 GPUs. The system follows a NUMA architecture with two sockets, where each processor is seated in its own socket and has two GPUs directly connected to it (Figure \ref{fig:numa_arch}). There are a total of 64 CPU cores, based on the Zen 3 architecture and a 2.7 GHz clock speed. The system's RAM capacity per node amounts to 768 GB, complemented by NVIDIA H100-94 SXM5 GPUs, each equipped with 94 GB of HBM2e memory, resulting in a total HBM2e memory capacity of 376 GB per node. The total GPU computational power across all four GPUs in each node is rated at 268 TFLOPS. The GPUs are interconnected via NVIDIA NVLink NV6. The system's topology includes PCIe interconnects classified as PIX, ensuring direct communication through a single PCIe bridge, reducing latency and CPU overhead \cite{NVIDIATopology}. The two CPU sockets are linked via AMD’s UPI, enabling direct memory access across NUMA domains.

The environment is running on Rocky Linux 9.3 with a kernel version of k08r01s02.novalocal (version 5.14.0-362.24.1.el9\_3.0.1) with CUDA 12.4, Nsight system version 2024.4.1 and SLURM version 23.11.6.

\subsection{Benchmarking software and procedures} 

\subsubsection{Implementation frameworks}
To perform the image recognition tests, a basic Distributed Data Parallel (DDP) module implementation \cite{DDP} based on PyTorch (version 2.2.2) and its torchvision (version 0.17.2) module for image processing was used. This framework provides a versatile and efficient toolkit, allowing for flexible workflow customization and integration of different techniques. However, during experiments, the NVIDIA's DALI library (version 1.43.0) based on CUDA 12.6 was also integrated and tested, which allowed offloading part of the data preprocessing workload from the CPU to the GPU. This approach helps reduce the CPU-GPU communication overhead during and between epochs, thereby improving the overall performance.
The fine-tuning of Large Language Models was conducted utilizing TorchTune (version 0.5.0) - a library within the PyTorch ecosystem. Torchtune uses recipes for single-device and distributed tuning, depending on whether one or more GPUs were used. These recipes can be subsequently modified using custom configuration files in order to perform a wide array of tests \cite{TorchtuneConfigs}. 

\subsubsection{Profiling tools}
NVIDIA Nsight Systems \cite{Nsight} is a comprehensive system-wide performance analysis tool designed to optimize and profile applications running on NVIDIA GPUs. It provides detailed insights into CPU and GPU interactions, memory usage, and the performance of CUDA kernels and APIs. By capturing a holistic view of the system's behaviour, nsys helps identify bottlenecks, analyse workload distribution, and optimize both CPU and GPU performance. Moreover, It is highly valuable for assessing the uniformity of load distribution across all GPUs and identifying bottlenecks in the algorithm's performance. A significant advantage of this tool is its minimal impact on the algorithm's execution time.

\subsection{Optimization techniques}
\subsubsection{Precision}
The investigated models support variable precision in computations. Double (64-bit), floating-point (32-bit), and half (16-bit) precision were used for the experiments. When reducing the precision of computations from double to float or half, several significant changes occur in the system's performance and resource utilization. The most immediate impact is a reduction in memory usage, as each value now occupies half the space. This reduction allows for larger batch sizes or higher resolution images to be processed simultaneously, enhancing the throughput of the model. Additionally, the reduced memory footprint decreases the data transfer load across the system's bus, particularly between the GPU and CPU, as well as across multiple GPUs, enabling faster communication and reducing potential bottlenecks. Moreover, the computational workload on the GPU decreases, as 32-bit operations require fewer computational resources than 64-bit operations, leading to increased processing speed. This can result in faster training and inference times, making the model more efficient overall. 
For half precision, the mixed precision technique was used. In this approach, half precision is used for most of the calculations to take advantage of its lower memory usage and faster computation, while float is used for operations requiring higher precision, such as maintaining the model's weights. By combining these two formats, mixed precision aims to reduce computational load and memory usage, thereby speeding up training and inference process. However, it is important to note that achieving such performance may come with a potential trade-off: reduced precision may lead to a slight decrease in numerical accuracy, which may impact the model's performance for some tasks, especially those that require precision.

\subsubsection{Pin\_memory}
During the execution of the algorithm, there is active communication and data exchange between the CPU and GPU. Among the investigated parameters $pin\_memory$ showed positive effects on the algorithm's performance. The parameter is attributed to \texttt{DataLoader} \cite{doi:10.2352/EI.2024.36.12.HPCI-196} and forces the system to use only page-locked memory and prevents intermediate data from being swapped to disk. By locking the memory pages in RAM, it allows for faster and more efficient data transfers between the CPU and GPU.

\subsubsection{Tensor structure}
In deep learning frameworks, data layout significantly affects computational efficiency. The two most common tensor formats are NCHW (Batch, Channels, Height, Width) and NHWC (Batch, Height, Width, Channels). The NCHW format is traditionally used in PyTorch and many deep learning models, as it aligns well with conventional convolution implementations. On the other hand, NHWC is optimized for modern GPU architectures, particularly those leveraging Tensor Cores, which perform matrix multiplications more efficiently when channels are stored in the last dimension. As a result, frameworks such as TensorFlow and NVIDIA DALI adopt NHWC to maximize hardware acceleration. The key advantage of NHWC over NCHW lies in memory access patterns and cache utilization. Modern GPUs rely on coalesced memory access, where threads in a CUDA warp (typically 32 threads) benefit from fetching adjacent memory addresses in a single request. In NCHW format, adjacent pixels in a channel are stored in separate memory locations, causing inefficient access patterns and increasing the number of memory transactions. Conversely, NHWC stores adjacent pixel values contiguously, allowing CUDA warps to load complete tensor segments efficiently, reducing global memory accesses and improving cache reuse. Additionally, NHWC optimizes L2 cache and shared memory usage by improving data locality. Since memory requests occur in larger, sequential blocks, NHWC reduces memory fragmentation, enabling GPUs to maximize bandwidth utilization. This leads to faster tensor transformations, improved convolution operations, and lower latency in deep learning pipelines. When comparing PyTorch DataLoader (default NCHW) to NVIDIA DALI (NHWC), the difference in cache efficiency explains why NHWC-based pipelines achieve higher throughput, particularly in multi-GPU training scenarios.

\subsection{Profiler metrics}
As a result of the tests, a comprehensive review of the metrics offered by profiling tools was conducted, which led to the selection of several key metrics. Table \ref{tab:metrics} presents selected metrics that have the greatest impact on the performance of the tested applications.

\begin{table}
\caption{Selected profiler metrics used for profiling} 
\label{tab:metrics}
%\adjustbox{max width=\linewidth}{
%\begin{tabular}{||c||}
\begin{tabular}{||m{13.5cm}||}
%\begin{xltabular}{\textwidth}{|X|}
\hline
%\centering 
\texttt{CUDA memcpy Host-to-Device}  \\
%\hline
This metric in NVIDIA Nsight Systems measures the time taken for data transfer from the CPU memory to the GPU memory. This metric is crucial for understanding the overhead associated with data movement in GPU-accelerated applications, particularly in scenarios where large datasets are transferred between the CPU and GPU. Analysing the \texttt{CUDA memcpy Host-to-Device} metric helps in identifying opportunities for optimization, such as using pinned memory (which can speed up transfers), optimizing the data size and transfer frequency, and overlapping data transfers with computation. \\
\hline
%\centering 
\texttt{cudaLaunchKernel} \\
%\hline
This metric measures the time taken to launch a GPU kernel from the host. This metric is essential for understanding the overhead involved in initiating the execution of computational tasks on the GPU.\\ 
\hline
%\centering 
\texttt{cudaStreamSynchronize} \\ 
%\hline
This metric is a function from the CUDA API designed to synchronize task execution on the GPU. It halts execution on the host (CPU) until all tasks in the specified stream on the GPU are completed. High values of this metric indicate a heavy load on the GPU, during which the CPU is in a waiting state. This can be an indirect indication of efficient GPU utilization. \\ 
\hline
%\centering 
\texttt{ncclDevKernel\_AllGather} \\ 
%\hline
This metric measures the time taken by the NCCL (NVIDIA Collective Communications Library) to perform the All-Gather operation using the Ring algorithm with the Low-Latency (LL) protocol. This metric is significant for analysing communication patterns in multi-GPU and distributed computing scenarios. High values in this metric indicate significant time spent on data communication between GPUs. In scenarios involving frequent or large-scale data sharing across GPUs, the performance of the All-Gather operation can become a critical factor influencing overall application speed and scalability. \\ 
\hline
%\centering 
\texttt{cudaEventSynchronize} \\
%\hline
The \texttt{cudaEventSynchronize} function is used to synchronize the host (CPU) thread with the GPU by forcing it to wait until all operations linked to a particular event are fully completed. This synchronization ensures that all GPU tasks associated with the event have finished, thus allowing precise control over the execution flow between the host and the device. This mechanism is particularly useful when subsequent operations on the CPU rely on the accurate completion of previous GPU computations, thereby providing a reliable barrier for coordinating concurrent processing tasks in heterogeneous computing environments. \\ 
\hline
%\centering 
\texttt{cudaEventDestroy} \\
%\hline
The \texttt{cudaEventDestroy} function is employed to properly release the memory and other resources associated with an event, once it is no longer needed. Efficient use of \texttt{cudaEventDestroy} helps to maintain optimal memory usage and prevent resource leaks, which is critical for applications requiring extensive GPU usage and resource-intensive parallel computations. By ensuring that unused events are correctly destroyed, this function plays a key role in managing the GPU’s resources effectively, thus enhancing the scalability and stability of high-performance parallel applications. \\ 
\hline
%\centering 
\texttt{cudaMemcpyAsync} \\ 
%\hline
This is an asynchronous memory copy function provided by CUDA that enables non-blocking data transfers between different memory spaces, such as from host (CPU) to device (GPU), device to host, or between device memories. This function is crucial for optimizing throughput in applications involving frequent data exchanges between the host and the device, reducing idle GPU time and improving overall performance. Additionally, \texttt{cudaMemcpyAsync} can be used in conjunction with pinned (page-locked) memory to further increase transfer speeds by leveraging direct memory access (DMA).  \\
\hline
%\end{xltabular}
\end{tabular}
%}
\end{table}

%%%%%%%%%%%%%%%%%%%%%%%%%%%%%%%%%%%%%%%%%%%%%%%%%%%%%%%%%%%%
\section{Profiling of image recognition}
\label{sec:cpu}

\subsection{Profiling methodology}

\paragraph{Benchmarking model}
The \texttt{MobileNet v2} \cite{Simonyan2014VeryDC} model was used to benchmark the image recognition process. It is a convolutional neural network with a depth of 53 layers, specifically designed for efficient performance in resource-constrained environments, such as mobile devices. The architecture is characterized by an inverted residual structure, wherein the input and output of each residual block are narrow bottleneck layers, while channel expansion occurs within the intermediate layers. This design contrasts with traditional residual models, where expanded representations are utilized at the input and output. \texttt{MobileNet v2} also employs depth-wise separable convolutions, significantly reducing the computational load and the number of parameters, while maintaining high accuracy. In addition, the removal of non-linearities in the narrow bottleneck layers prevents information loss, enhancing the model's capacity for generalization. These architectural choices make \texttt{MobileNet v2} highly suitable for deployment in environments with limited computational resources. The model also supports adjustable width multipliers, allowing for scalability depending on the available computational power. The implementation of \texttt{MobileNet v2} across several GPUs results in significant enhancements in both training and inference speeds, especially when processing extensive datasets. Its lightweight design and efficient use of parameters allow the model to scale proficiently with parallel computing, thereby maximizing the performance of multi-GPU configurations.

\paragraph{Workload distribution strategies}
In a series of tests, a number of data distribution methods were analysed to determine the most effective in terms of utilizing GPU resources in large-scale machine learning projects. The most efficient strategy was the Distributed Data Parallelism with Distributed Sampler (DDP-DS) which manages dividing the dataset into samples and parallelized workload distribution. This necessitates the relocation of Data Loader logic into a parallelized process function. Nevertheless, this adjustment ensures that only the necessary data segments are transmitted to the GPUs. DDP-DS effectively reduces the frequency of data transfers from the host to the device, as each process is allocated a distinct subset of data, thereby minimizing unnecessary data movement. Based on profiler metrics, the \texttt{cudaMemcpyHostToDevice} operations are both quicker and less frequent in DDP-DS, leading to a reduced overall duration for these operations.

\paragraph{Examined dataset}
To evaluate different approaches to GPU load distribution, it is crucial to ensure that all interventions do not affect the algorithm's outcomes. This requires stable input data and a quantifiable metric for neural network training quality. \texttt{MNIST} \cite{MNIST} is a dataset containing monochrome handwritten digit images, each sized 28x28 pixels, with 60,000 training images and 10,000 test images. This change enables to evaluate the training quality and ensure that load distribution mechanisms do not interfere with the algorithm's logic. Additionally, it was needed to convert the monochrome images to "coloured" images, as tested models work with 3-channel images \cite{MobileNetParams}. Furthermore, following the completion of multiple tests, it became evident that the dimensions of 28x28 pixels were insufficient to effectively engage the GPU, as tasks associated with memory management and communication were prioritized over computational processes. Therefore, the dimensions of the analysis output images were modified from 100x100 to 500x500 pixels in order to evaluate the impact of this alteration on the GPU load.

\paragraph{Model training quality}
In the research undertaken, both the effects of different parameters on the model's runtime and their impact on computational accuracy were analysed. In summary, the key parameter for achieving stable and acceptable results was the number of training epochs. The goal of testing was to achieve 98-99\% accuracy on the test dataset within 20 training epochs. Increasing the image size resulted in longer model execution time but more accurate results.

\paragraph{Scenarios}
In order to test various common image processing approaches for their native resolution, two scenarios were defined:

\textbf{Scenario A} - due to the need to impose a significant computational load on the GPU, in this scenario, upsampling of test images was performed. It consists in increasing the spatial resolution of an image or a feature map by generating additional pixel values. In the context of AI and deep learning, upsampling is commonly used in tasks such as image segmentation, super-resolution, and generative modelling \cite{ronneberger2015unetconvolutionalnetworksbiomedical, kundu2020attentionbasedimageupsampling, vasconcelos2022cufcontinuousupsamplingfilters}. Additionally, it was necessary to perform channel duplication to match the RGB format and normalization. Scenario A was used under precision and $pin\_memory$ testing.

\textbf{Scenario B} - in this approach, upsampling was eliminated by using 300x300px colour images (~3GB/set) and the only preprocessing operation required was normalization. Considering that the dataset size after unpacking into tensors expands from 3GB to about 68GB, this gives a total memory footprint of 272GB using four GPUs (each process keeps its own copy of the data). Due to memory constraints on GPU cards, 300x300px resolution turned out to be the maximum feasible. Scenario B focuses on comparing PyTorch DataLoader with NVIDIA DALI and optimizing the tensors.

\subsection{Performance and optimization evaluation}
\begin{figure}[h]
    % \begin{sidewaysfigure}
        % \centering
        \includegraphics[width=0.49\linewidth]{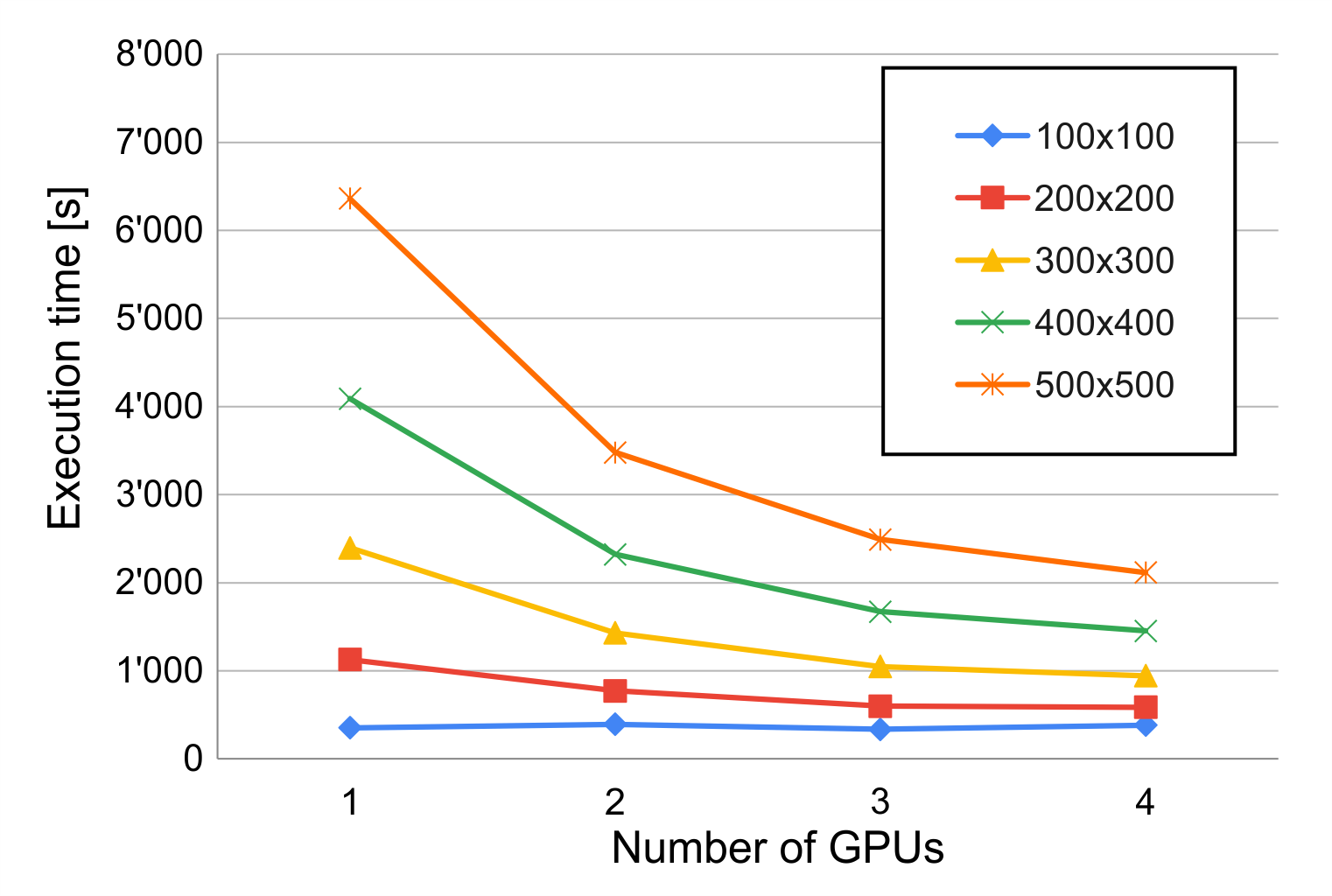}
        \includegraphics[width=0.49\linewidth]{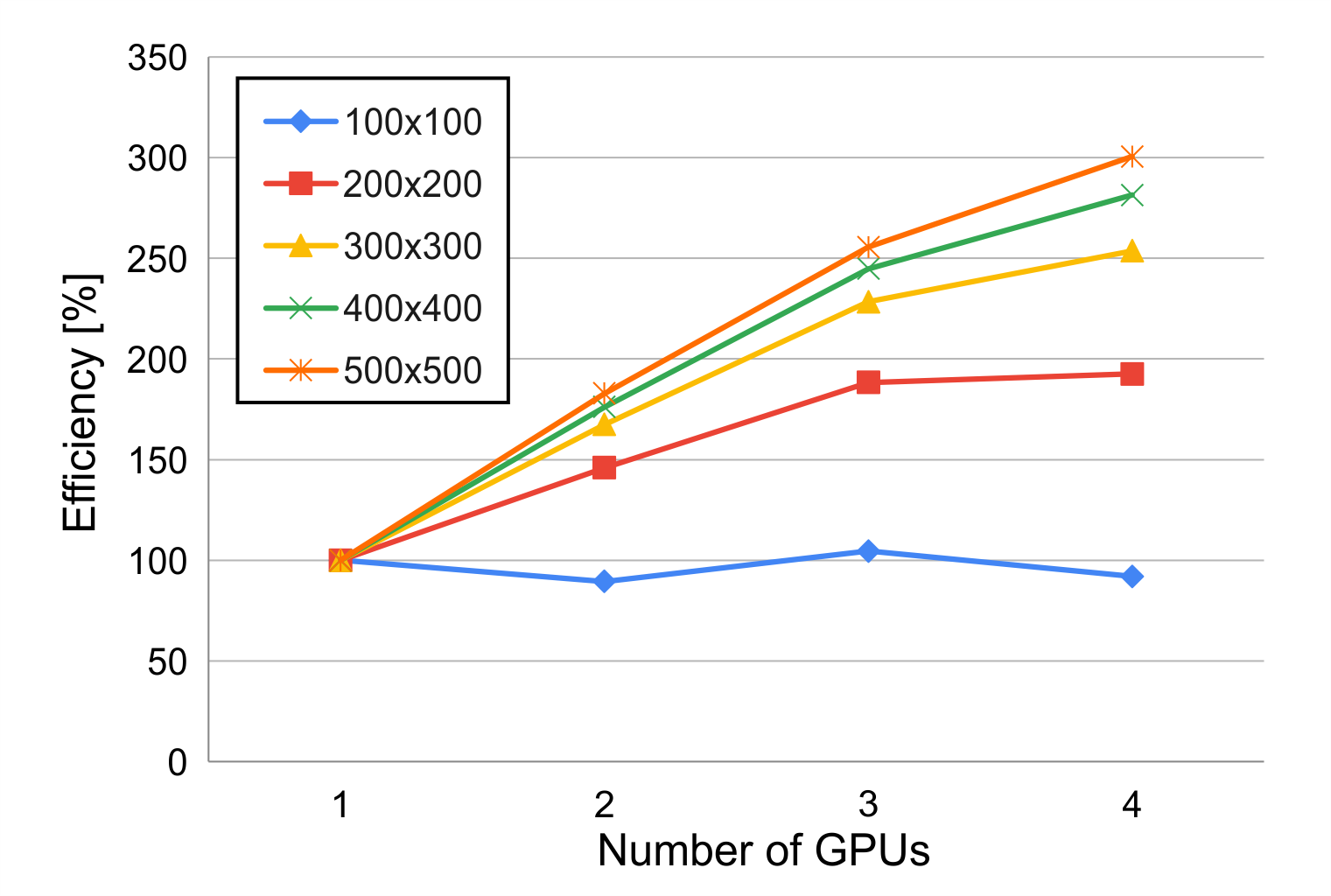}
    % \end{sidewaysfigure}
    \caption{Execution time (left) and efficiency (right) graphs for image sizes from 100x100 to 500x500, \texttt{FP64}, $pin\_memory = false$}
    \label{fig:diff_sizes_graph}
\end{figure}

The Figure \ref{fig:diff_sizes_graph} presents the performance results of the DDP-DS algorithm for image sizes ranging from 100x100 pixels to 500x500 pixels. From these graphs, it is evident that the benefits of adding new GPUs become noticeable only at a certain image size, which can create sufficient load on the computational power of the GPUs. The graph shows that effective utilization of 4 GPUs is observed with image sizes starting from 300x300 pixels.

\subsubsection{Optimization results}

\paragraph{Precision}
Switching from double (\texttt{FP64}) to float (\texttt{FP32}) or half (\texttt{FP16}) precision yields a significant improvement in all presented cases. In table \ref{tab:precision_table} it can be observed a consistent increase in computational speed by 54-68\% for float and 110-152\% for half compared to using the same parameter set with double precision. Additionally, as shown in the 'Success test dataset' column, the quality of results does not deteriorate when transitioning from double to float or half precision. This can be attributed to the fact that training is performed over 20 epochs, during which the network achieves satisfactory training quality.

\begin{table}[h]
    \centering
    \adjustbox{max width=\linewidth}{
    \begin{tabular}{c c c c c c}
        \toprule
        \textbf{GPUs} & \textbf{Precision} & \textbf{Total} \textbf{time [s]} & \textbf{Inference} \textbf{accuracy} \textbf{[\%]} & \textbf{Efficiency} \textbf{compare} \textbf{to 1 GPU} & \textbf{Efficiency} \textbf{compare} \textbf{to \texttt{FP64}} \\
        \midrule
        1 & FP64 & 6\,266 & 99.43 & 100\% & -       \\
        1 & FP32  & 3\,721 & 99.32 & 100\% & 168.38\% \\
        1 & FP16  & 2\,477 & 99.57 & 100\% & 252.93\% \\
        2 & FP64 & 3\,339 & 99.20 & 187.66\% & -       \\
        2 & FP32  & 2\,094 & 99.24 & 177.69\% & 159.45\% \\
        2 & FP16  & 1\,445 & 97.94 & 171.41\% & 231.07\% \\
        3 & FP64 & 2\,318 & 98.70 & 270.31\% & -       \\
        3 & FP32  & 1\,483 & 99.03 & 250.91\% & 156.30\% \\
        3 & FP16  & 1\,049 & 99.21 & 236.12\% & 220.97\% \\
        4 & FP64 & 1\,937 & 99.32 & 323.48\% & -       \\
        4 & FP32  & 1\,256 & 98.88 & 296.25\% & 154.21\% \\
        4 & FP16  & 919 & 99.20 & 269.53\% & 210.77\% \\
        \bottomrule
    \end{tabular}}
    \caption{Comparison of algorithm performance for \texttt{FP16}, \texttt{FP32} and \texttt{FP64}, $pin\_memory=FALSE$}
    \label{tab:precision_table}
\end{table}

% https://docs.nvidia.com/deeplearning/performance/mixed-precision-training/index.html - tłumaczenie mechanizmu mixed precision
Upon examining the profiler and comparing changes in various metrics, it can be observed that the largest reduction in time occurs at the [CUDA memcpy Host-to-Device] stage and \texttt{cudaMemcopyAsync}. The Host-to-Device copy time decreases proportionally with a twofold reduction in precision for float, and a fourfold reduction for half precision, from 226s for \texttt{FP64} to 104s for \texttt{FP32} and 50s for \texttt{FP16}. \texttt{cudaMemcopyAsync} also decreases proportionally: 1\,334s for \texttt{FP64}, 724s for \texttt{FP32} and 478s for \texttt{FP16}. This is the most significant change. Among other metrics, there are also changes in $ncclDevKernel\_Broadcast$, where the time decreases proportionally from 40s for \texttt{FP64} to 27s for \texttt{FP32} and 15s for \texttt{FP16}. Among the remaining metrics, it is also worth noting the \texttt{cudaEventDestroy} metric, whose execution time decreases threefold or even more, rather than twofold, with each reduction in precision, from 205s for \texttt{FP64} to 84s for \texttt{FP32} and 16s for \texttt{FP16}.

\paragraph{Pin\_memory}
Table \ref{tab:pinMemory_table} presents a performance improvement from using $pin\_memory$, ranging between 16\% and 30\%, with the efficiency decreasing as the number of GPUs increases. The analysis shows that switching to lower precision (\texttt{FP32} and \texttt{FP16}) from double precision (\texttt{FP64}) provides significant performance improvements, with up to 210\% speed increase for \texttt{FP16} without sacrificing inference accuracy. Using pinned memory ($pin\_memory$) further enhances performance by 16-30\%, though the efficiency gain decreases as the number of GPUs increases. Profiling reveals that the most substantial time reduction occurs during CUDA Host-to-Device transfers and \texttt{cudaMemcopyAsync}, with \texttt{FP16} achieving the lowest times. However, increasing the number of GPUs and enabling $pin\_memory$ also leads to increased execution times for NCCL kernels, diminishing some of the benefits obtained from faster \texttt{Host-to-Device} communication. Overall, reduced precision and optimized memory handling are crucial in enhancing computational efficiency in multi-GPU environment.

\begin{table}[h]
    \centering
    \adjustbox{max width=\linewidth}{
    \begin{tabular}{c c c c c c}
        \toprule
        \textbf{GPUs} & \textbf{Pin} \textbf{memory} & \textbf{Total} \textbf{time [s]} & \textbf{Inference} \textbf{accuracy} \textbf{[\%]} & \textbf{Efficiency} \textbf{compare} \textbf{to 1 GPU}  & \textbf{Efficiency} \textbf{compare} \textbf{to $FALSE$} \\
        \midrule
        1 & FALSE & 3\,792 & 99.30 & 100.00\% & -       \\
        1 & TRUE  & 2\,921 & 99.41 & 100.00\% & 129.78\% \\
        2 & FALSE & 2\,205 & 99.24 & 171.93\% & -       \\
        2 & TRUE  & 1\,733 & 99.48 & 168.56\% & 127.23\% \\
        3 & FALSE & 1\,579 & 99.30 & 240.01\% & -       \\
        3 & TRUE  & 1\,283 & 98.70 & 227.57\% & 123.05\% \\
        4 & FALSE & 1\,358 & 99.20 & 279.08\% & -       \\
        4 & TRUE  & 1\,165 & 99.36 & 250.79\% & 116.62\% \\
        \bottomrule
    \end{tabular}}
    \caption{Comparison of algorithm performance for $pin\_memory$ flag, precision \texttt{FP32}}
    \label{tab:pinMemory_table}
\end{table}

\subsubsection{Metrics analysis}
\begin{figure}[h]
    \centering
    \includegraphics[width=0.65\linewidth]{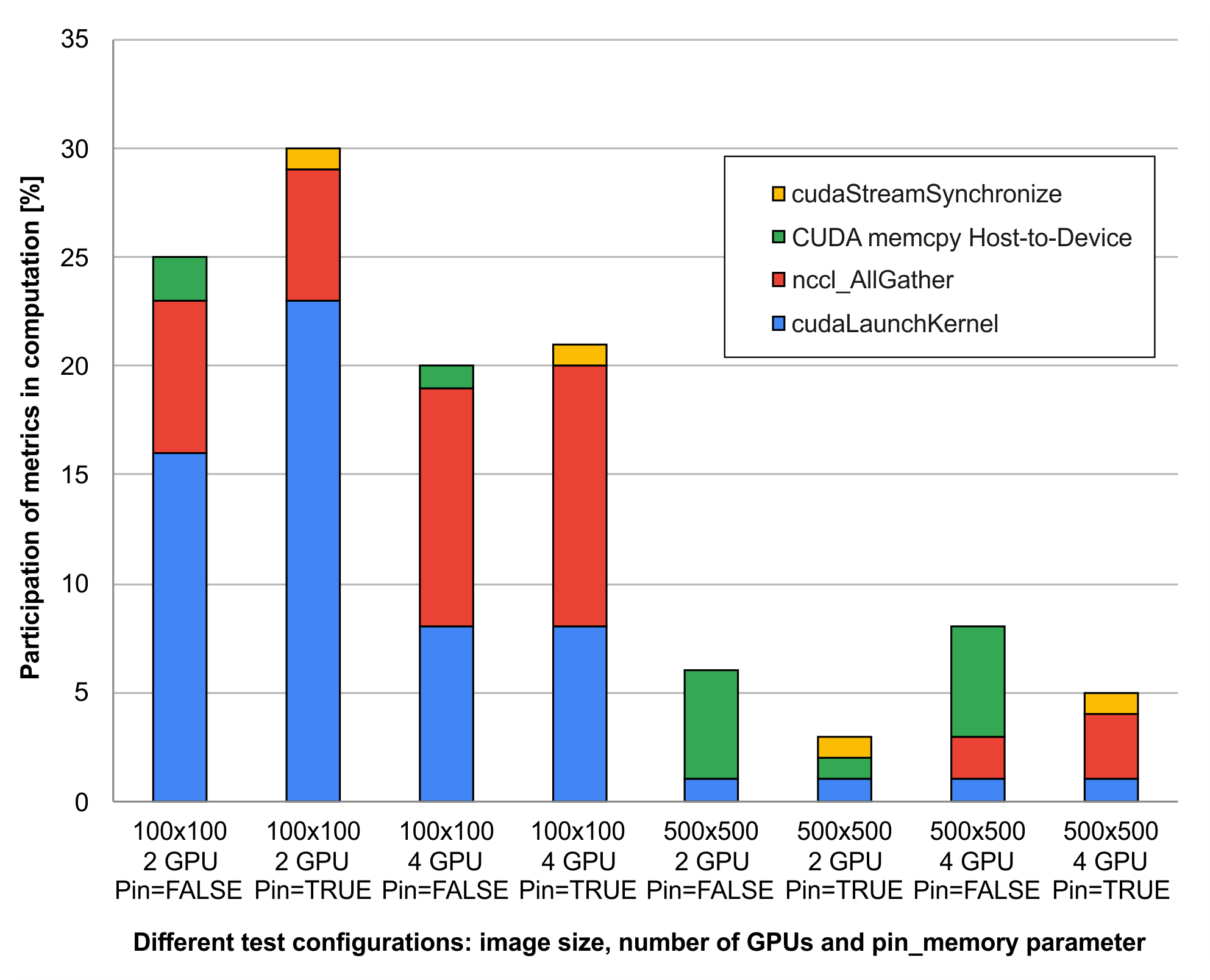}
    \caption{Diagram with four metrics times from NSight System presented in percent}
    \label{fig:metric_percent_diagram}
\end{figure}

\begin{figure}[h]
    \centering
    \includegraphics[width=0.65\linewidth]{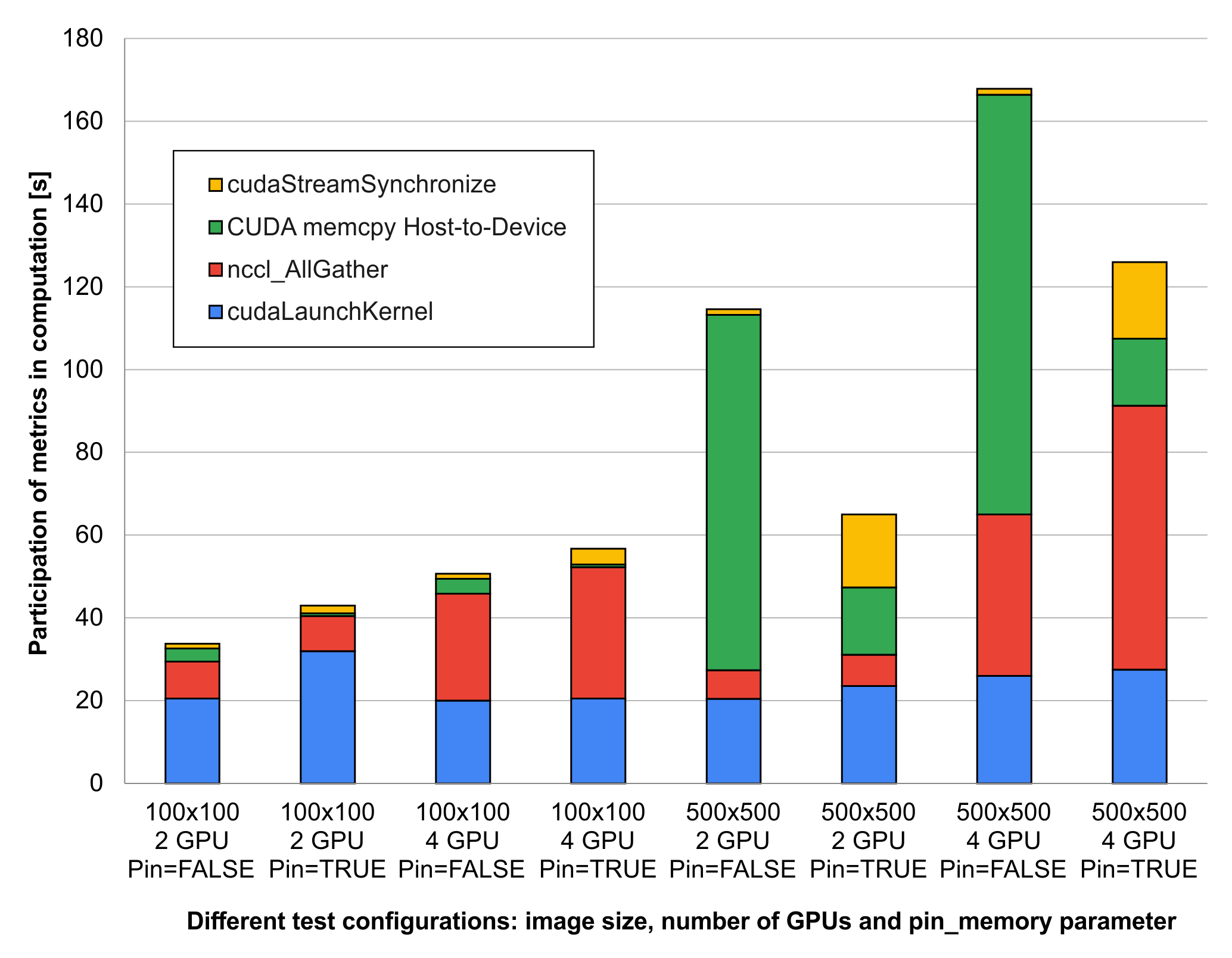}
    \caption{Share of metrics in the calculation time considering different test configurations including: image size, number of GPU and $pin\_memory$ parameter}
    \label{fig:metric_sec_diagram}
\end{figure}

While analysing the data provided by the profiler, it was found that these optimization techniques affect the execution time of different kernels and memory operations. Four of the most interesting metrics were selected: \texttt{CUDA memcpy Host-to-Device}, \texttt{cudaLaunchKernel}, \texttt{cudaStreamSynchronize}, and \texttt{ncclDevKernel\_AllGather}. Figure \ref{fig:metric_percent_diagram} and Figure \ref{fig:metric_sec_diagram}, present 8 tests for \texttt{FP32} with 3 input parameter variations: image size (100x100 or 500x500), number of GPUs (2 or 4), and $pin\_memory$ (true or false).

In Figure \ref{fig:metric_percent_diagram} diagram, it can be observed that \texttt{cudaLaunchKernel} and \texttt{ncclDevKernel\_AllGather\_RING\_LL} take relatively more time when the image size is small. This indicates that the algorithm spends more time on computation management and distribution than on the computations themselves. Additionally, it can be seen that the time for \texttt{CUDA memcpy Host-to-Device} significantly decreases when $pin\_memory$ is enabled, but the time spent on \texttt{cudaStreamSynchronize} increases.

Figure \ref{fig:metric_sec_diagram}, presents the same data, but with time measured in seconds. A significant reduction in \texttt{CUDA memcpy Host-to-Device} time is observed from 85-101s to 16s for 500x500px image size, and from 3.1-3.6s to 0.66s for 100x100px image size. Metric \texttt{cudaStreamSynchronize} shows increasing time from 1.3s to 18s for 500x500px image size. This can likely be explained by the fact that with such an acceleration in \texttt{HtoD} communication, various bottlenecks are starting to be noticed, causing threads to spend more time waiting during synchronization. It can also be observed that with an increase in the number of GPUs, the execution time of NCCL kernels increases from 7 to 39 seconds. Moreover, if $pin\_memory$ is enabled, the execution time rises to 63 seconds, significantly reducing the advantage gained from savings on \texttt{HtoD} operations.

% Wnioski
The analysis highlights the complexities of balancing optimizations for different aspects of GPU execution. While optimizations such as enabling $pin\_memory$ can greatly reduce the time spent on memory transfers, they may introduce other bottlenecks, such as increased synchronization overhead. Additionally, the number of GPUs affects performance non-linearly, with NCCL communication overhead increasing significantly as more GPUs are added, especially when using $pin\_memory$.

\subsubsection{Loss function}
\begin{figure}[h]
    \centering  
    \includegraphics[width=0.6\linewidth]{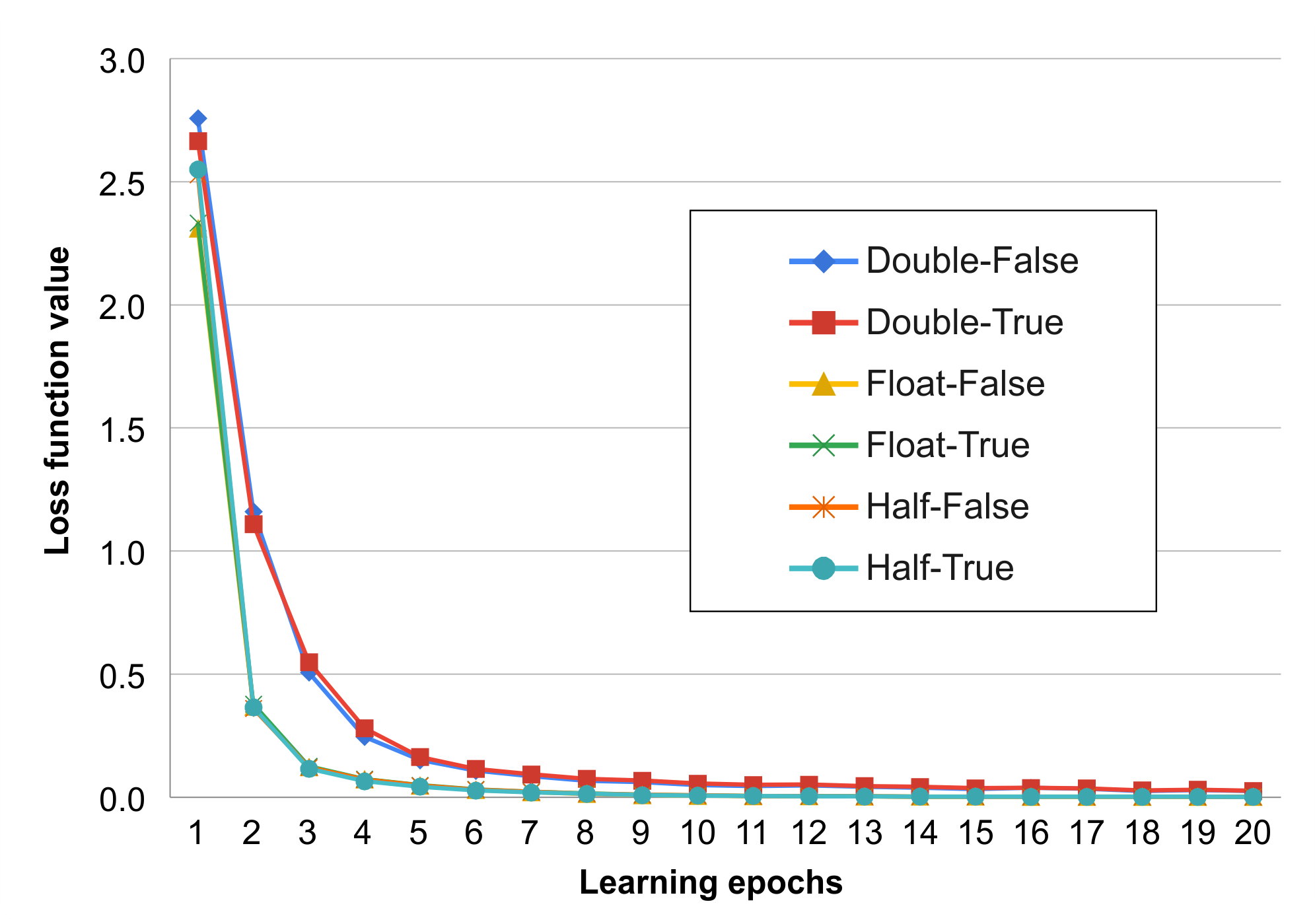}
    \caption{Loss curve for 6 tests with different precision (double, float, half) and $pin\_memory$ (TRUE, FALSE)}
    \label{fig:loss_curve_diagram}
\end{figure}

The MNIST dataset is a relatively simple task for the MobileNetV2 model, and in this study, it is evident that the model can achieve satisfactory inference accuracy after just 2-3 epochs of training. The simplicity of the dataset allows for rapid convergence, which makes it an ideal benchmark for comparing different precisions and data transfer optimizations in training deep learning models.
The Figure \ref{fig:loss_curve_diagram} presents error curves for training over 20 epochs reveal that the use of double precision results in a slower convergence compared to float and half precisions. Both float and half precisions yield similar loss rates throughout the training process, indicating that the reduced precision of half is sufficient to achieve comparable model accuracy. The slightly delayed convergence of double precision suggests that increased numerical precision does not provide a substantial benefit and might, in fact, lead to reduced efficiency due to increased computational demands.

In addition, training with half precision shows a notable improvement in terms of computational speed compared to float and double precisions, making it a more efficient choice without sacrificing accuracy. The analysis also indicates that the use of pinned memory ($pin\_memory$=True) does not have a significant impact on the training loss curves, suggesting that its primary effect lies in optimizing data transfer efficiency between CPU and GPU, rather than directly influencing the learning dynamics or the model's convergence behaviour.

% Wnioski
In summary, the results indicate that reducing precision to float or half precision can significantly improve computational speed without compromising accuracy, or even improving time to achieve minimal loss, making it a viable strategy for optimizing training efficiency. Moreover, while pinned memory helps improve data transfer performance, its effect on overall training convergence is minimal, highlighting that its primary benefit lies in improving data handling efficiency rather than affecting learning dynamics.

\subsection{Evaluation of findings for image recognition}
In this section, the focus is placed on comparing the PyTorch DataLoader and NVIDIA DALI, analysing their architectural differences, and evaluating NSYS reports according to Scenario B.

\subsubsection{DDP perfomance}
\begin{figure}[h]
% https://docs.google.com/spreadsheets/d/1qnZ6fbKqZ1i-WrTSuTHonoh7Ea7qwuUpoohJNNETdXc/edit?gid=0#gid=0
% Arkusz - Graphs and Tables 1
    % \centering
    \includegraphics[width=0.49\linewidth]{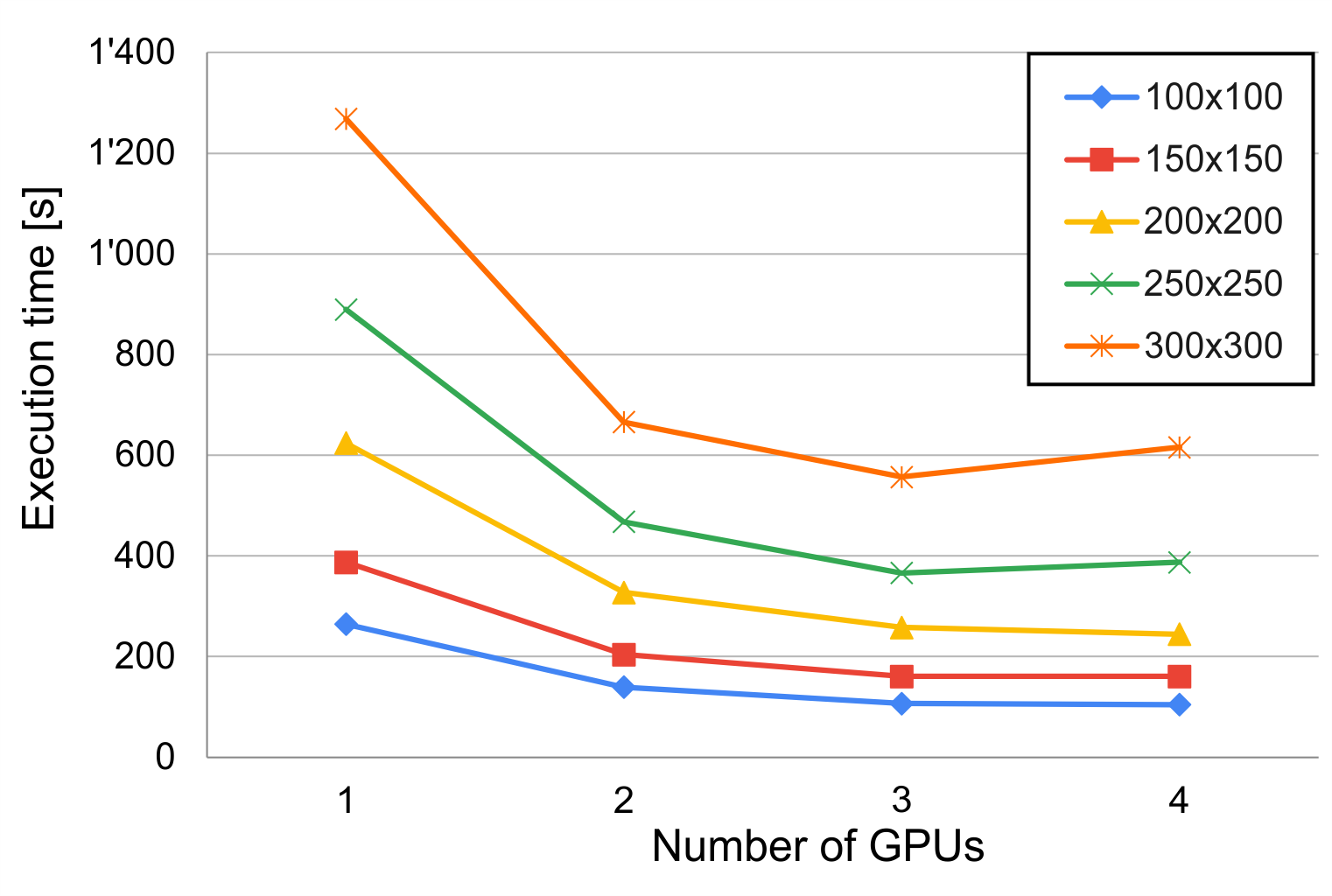}
    \includegraphics[width=0.49\linewidth]{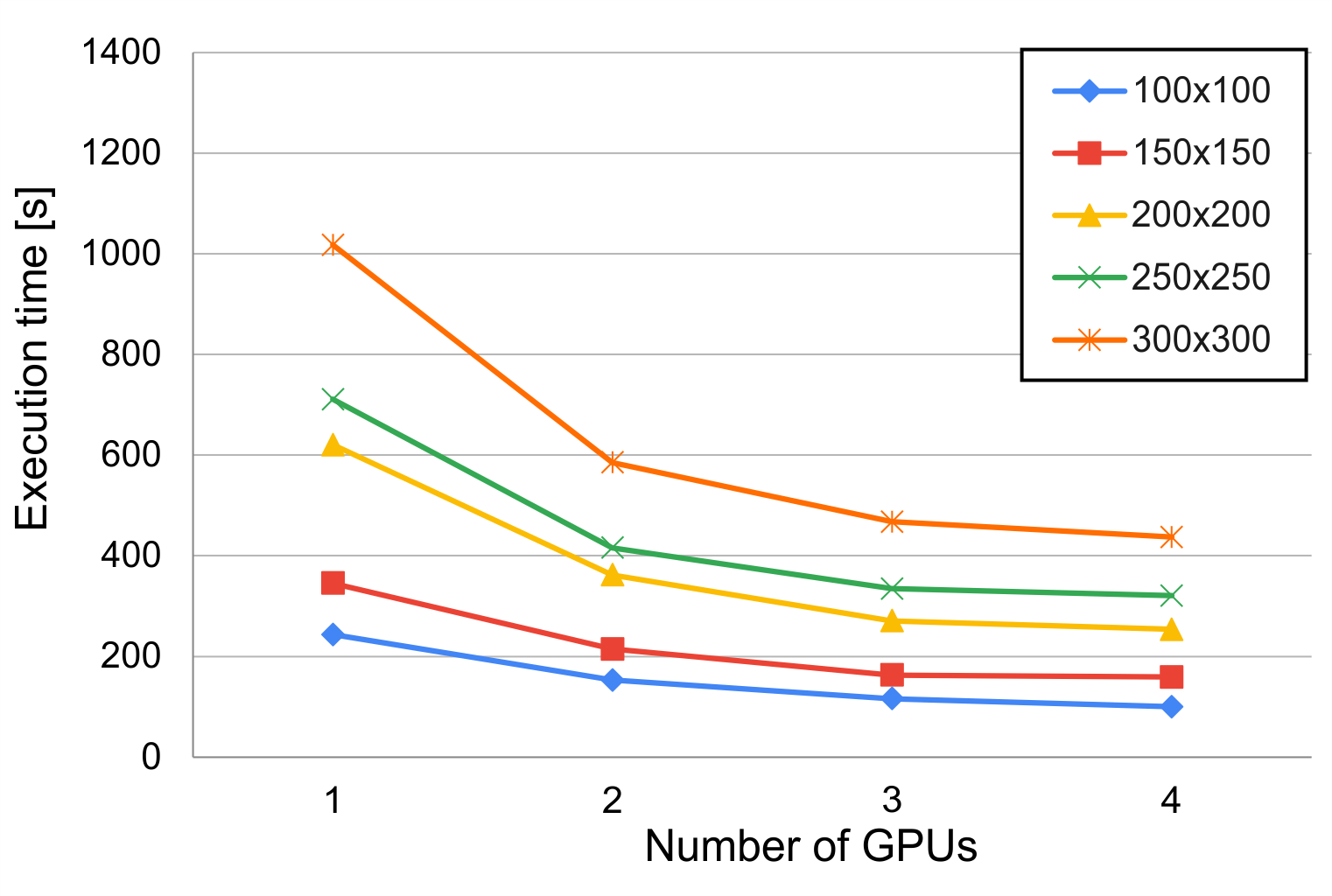}
    \caption{Execution time for DataLoader (left) and DALI (right) graphs for image sizes from 100x100 to 300x300, \texttt{FP32}, $pin\_memory = true$}
    \label{fig:dl_vs_dali_time_graph}
\end{figure}

% Opis zmian związanych z scenariuszem B: więcej danych do transferu. Opisać problem z błędną alokacją pamięci dla DL i spad wydajności na 3-4 GPU
The key difference between Scenario A and Scenario B for PyTorch DataLoader is that the dataset on disk occupies gigabytes (Scenario B) rather than megabytes (Scenario A), which increases the time required to load the dataset from disk into RAM. Since the dataset is stored on HDD, the loading process takes a significant portion of the script's runtime, typically around 35 seconds for 300×300px dataset. No other major changes occur. The volume of CPU-GPU data transfer for DataLoader does not increase, as DataLoader performs pre-processing (including upsampling) on the CPU, meaning that only the already upsampled images are sent to the GPU.
However, as presented on Figure \ref{fig:dl_vs_dali_time_graph}, when transitioning from 2 GPUs to 3-4 GPUs, the DataLoader exhibits a significant performance drop, slowing down the entire process. DALI, on the other hand, does not suffer from this issue.

\begin{figure}[h]
    % \begin{sidewaysfigure}
        % \centering
        \includegraphics[width=0.99\linewidth]{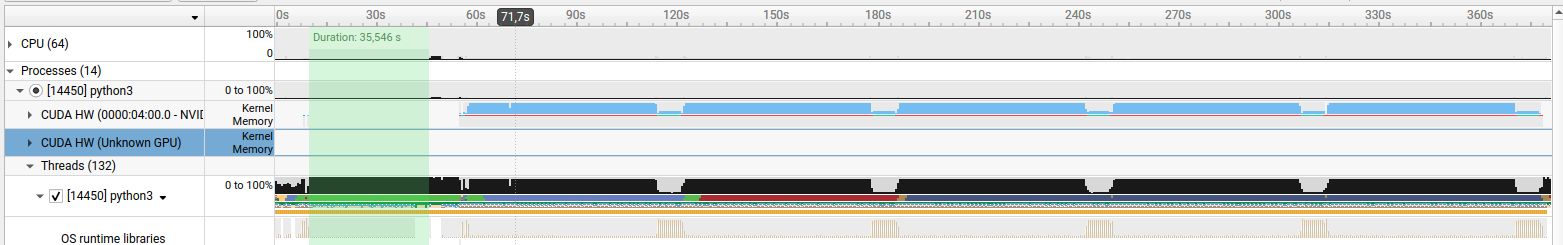}
        \includegraphics[width=0.99\linewidth]{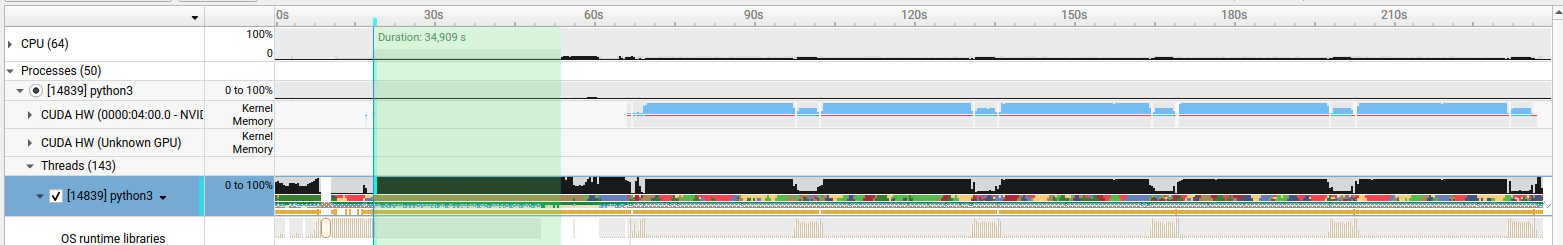}
        \includegraphics[width=0.99\linewidth]{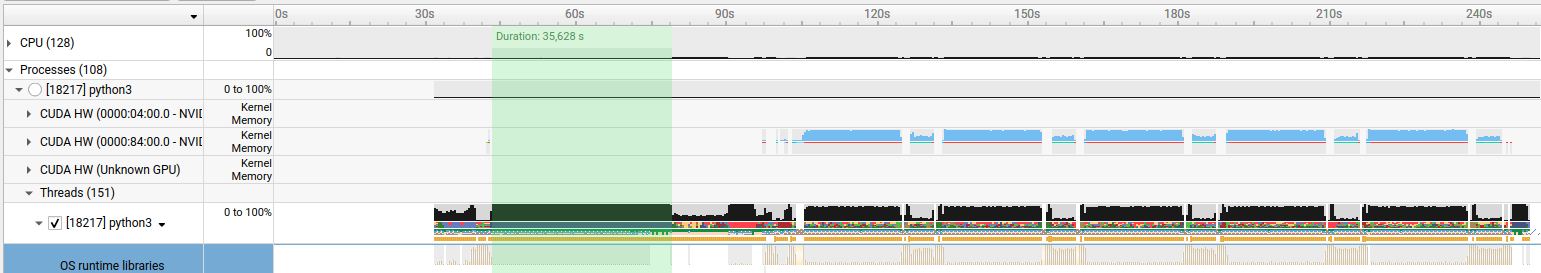}
        \includegraphics[width=0.99\linewidth]{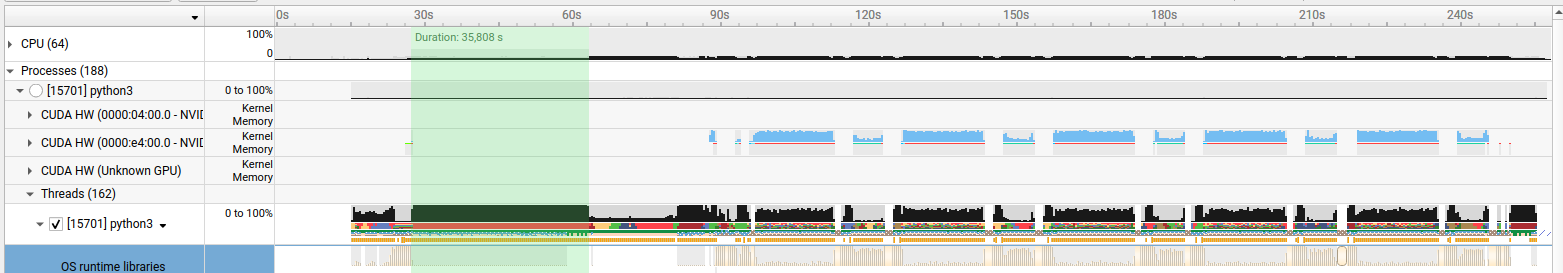}
    % \end{sidewaysfigure}
    \caption{NSYS time graph reports for 1 GPU(up), 2 GPU, 3 GPU, 4 GPU(down) DataLoader with picture resolution 300x300, \texttt{FP32}, $pin\_memory = true$}
    \label{fig:dl_time_graph}
\end{figure}

Investigating the NSYS report (Figure \ref{fig:dl_time_graph}), it can be observed that the training process for DataLoader remains largely unchanged when transitioning from 1 to 2 GPUs. However, with 3-4 GPUs, it becomes evident that pauses appear, likely due to increased inter-socket memory access overhead, before each training epoch and inference phase, significantly impacting the overall model training time. The hypothesis is that these pauses are caused by memory for processes on the second socket being allocated to the first socket, forcing frequent data transfers between sockets via UPI, which causes additional latency and performance degradation. These pauses are likely the reason why DataLoader starts losing performance when using 4 GPUs.

\begin{figure}[h]
    % \begin{sidewaysfigure}
        % \centering
        \includegraphics[width=0.99\linewidth]{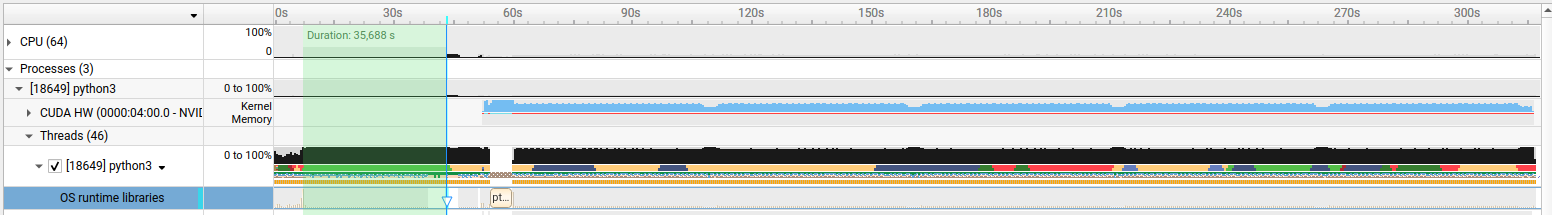}
        \includegraphics[width=0.99\linewidth]{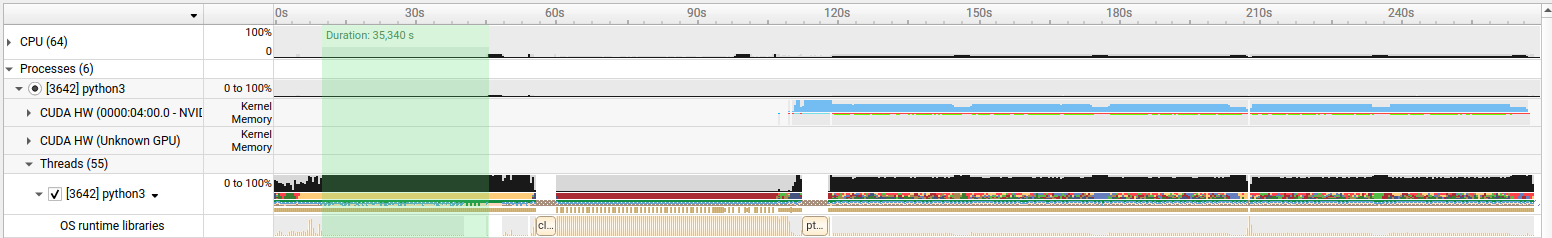}
        \includegraphics[width=0.99\linewidth]{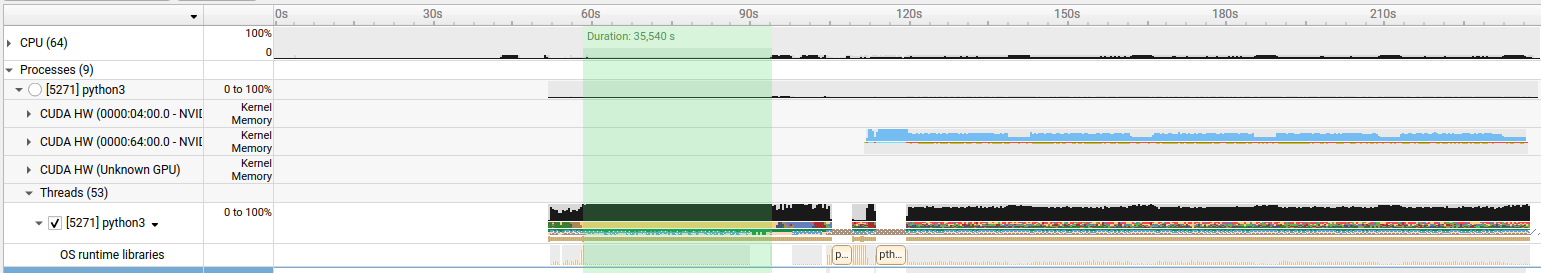}
        \includegraphics[width=0.99\linewidth]{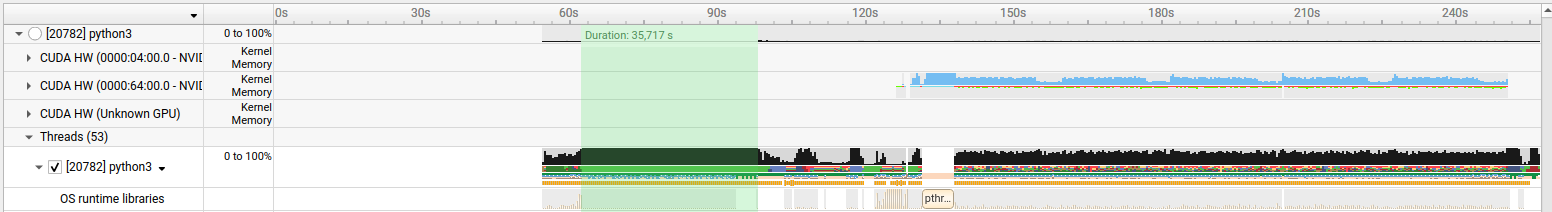}
    % \end{sidewaysfigure}
    \caption{NSYS time graphs reports for 1 GPU (up), 2 GPU, 3 GPU, 4 GPU (down) DALI with picture resolution 300x300, \texttt{FP32}}
    \label{fig:dali_time_graph}
\end{figure}

The comparison of the NSYS report for DALI (Figure \ref{fig:dali_time_graph}) shows that utilization of GPU remains consistent, with no pauses between epochs, even when running on 4 GPUs. This suggests that the transition from Scenario A to Scenario B itself does not cause the issue, indicating that the problem lies within DataLoader. One possible explanation is that DataLoader does not explicitly optimize memory locality in NUMA architectures, which could lead to processes running on GPUs attached to the second NUMA socket accessing memory allocated on the first socket. This may introduce frequent cross-socket memory transfers over UPI, increasing latency and reducing overall performance.

It is worth mentioning that the study tested scenario B with images larger than 300×300px. For 500×500px images, the dataset occupied around 200 GB of RAM, meaning that memory allocation for 2 GPUs could not fit within a single NUMA socket RAM, preventing from running DataLoader with 2 or 4 GPUs. However, DALI was successfully launched on 2 GPUs, further pointing to potential memory allocation issues within DataLoader. Furthermore, it is noteworthy that the initialization processes of DataLoader and DALI exhibit distinct differences. DataLoader initializes all threads prior to commencing parallel data reading. Conversely, DALI begins by reading metadata and initial data with a single thread, subsequently initializing the other threads, which then proceed to read the data. As a result, in Figure \ref{fig:dali_time_graph}, gaps can be perceived in the graph for 2 and 4 GPUs.

\subsubsection{Tensor structure NHWC}
\begin{figure}[h]
% https://docs.google.com/spreadsheets/d/1qnZ6fbKqZ1i-WrTSuTHonoh7Ea7qwuUpoohJNNETdXc/edit?gid=0#gid=0
% Arkusz - Graphs and Tables 1
    \centering
    \includegraphics[width=0.65\linewidth]{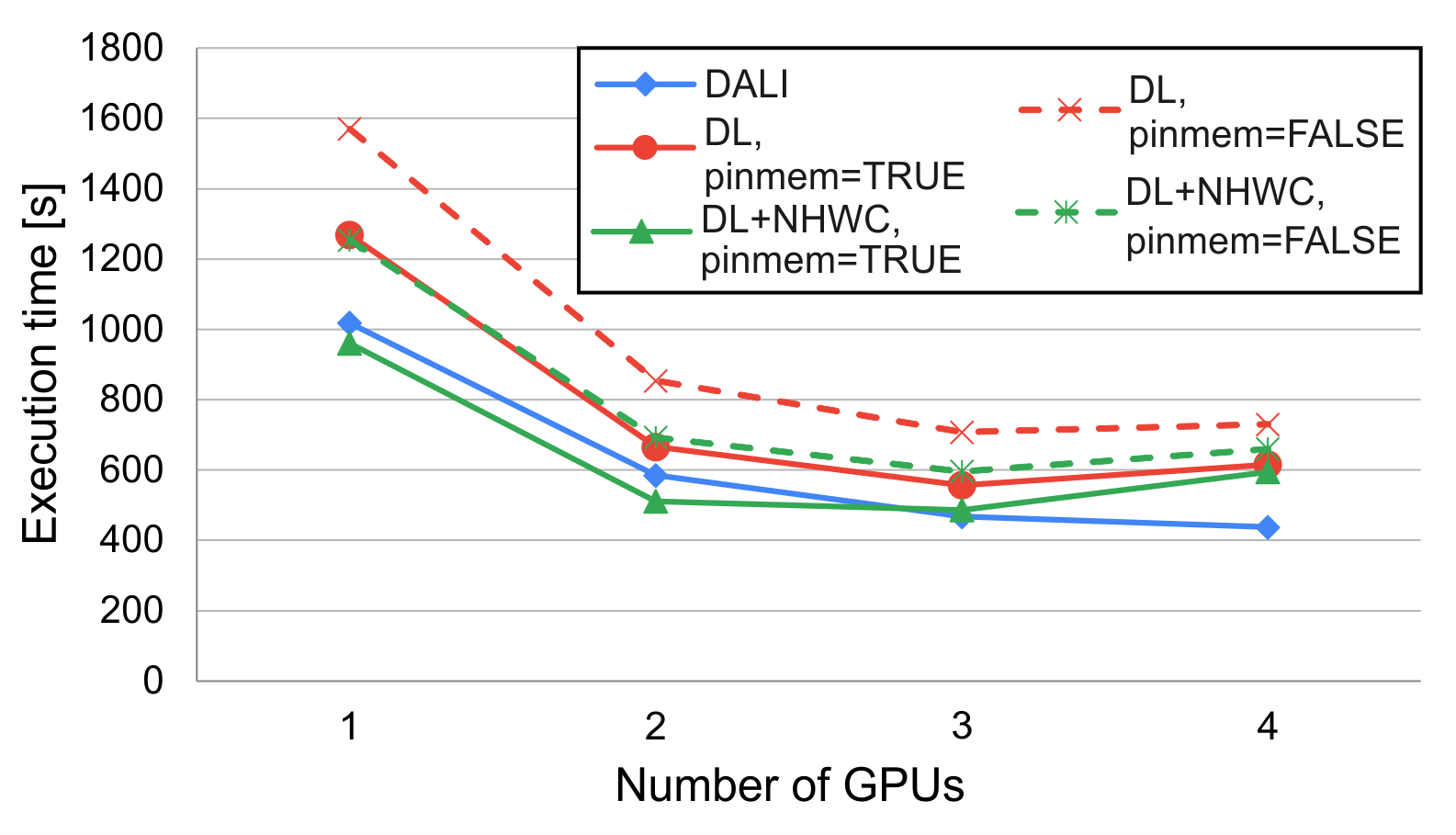}
    \caption{Comparison of execution time for DALI, DataLoader and DataLoader with NHWC tensor structure for image sizes 300x300, \texttt{FP32}, $pin\_memory = true$}
    \label{fig:dl_nhwc_vs_dali_compare_graph}
\end{figure}

During the comparison of NSYS reports for DataLoader and DALI, it was discovered that DataLoader, by default, uses the NCHW tensor structure, whereas DALI utilizes NHWC. This is a crucial difference that explains DALI’s advantage over DataLoader when running on a single GPU under otherwise identical conditions.
Figure \ref{fig:dl_nhwc_vs_dali_compare_graph}, presents that after switching DataLoader’s tensor structure to NHWC, the performance for 1 and 2 GPUs becomes very close to that of DALI, and the structure of the tables in the NSYS reports becomes almost identical. This results in more efficient cache utilization, making DataLoader and DALI nearly equivalent in performance for 1-2 GPUs. However, the issue of performance degradation for 3-4 GPUs remains, further indicating that the problem lies within DDP's interaction with DataLoader.

%%%%%%%%%%%%%%%%%%%%%%%%%%%%%%%%%%%%%%%%%%%%%%%%%%%%%%%%%%%%
\section{Profiling of Large Language Models}

In this part of the study, the performance of LLAMA3-8B LoRA \cite{10695440} fine-tuning process using differing datasets in a multi-GPU environment was evaluated. Additionally, data transfer/communication operations are also explored in this context. Finally, a comparison of these approaches and conclusions are provided.

\subsection{Model specification}
LLama3 is a large language model developed by Meta trained to perform tasks such as text generation, translation, question answering, and sentiment analysis \cite{minaee2024largelanguagemodelssurvey}. It utilizes transformer architecture with Grouped-Query Attention \cite{ainslie2023gqatraininggeneralizedmultiquery} to improve inference efficiency. Training was performed on the internet-scale corpus of diverse texts of over 15T (trillion) tokens. It uses a tokenizer with a vocabulary of 128K tokens, and its context length (the number of tokens considered by the model to predict the next word) is set at 8,192 tokens \cite{dubey2024llama3herdmodels}. The LLAMA3-8B variant of Llama consists of 8 billion parameters, with March, 2023 knowledge cut-off date. Smaller size makes it less memory demanding and easier to deploy in various applications. The model can be fine-tuned on specific tasks by further training it on task-specific data, allowing it to perform better on those tasks.

\subsection{LLM fine-tuning process}
Fine-tuning is a process separate from model training, that can help the model improve its performance on particular tasks where additional knowledge is required beyond what can be learned from general text data alone.

\subsubsection{Tuning techniques}
There are several approaches of LLMs tuning. Because of large number of parameters present in these models, efficient approaches are necessary. One example is LoRA (Low-Rank Adaptation) - a technique involving attaching a small number of low-rank adapter layers to a pre-trained LLM. These are trained on specific downstream tasks and improve performance without modifying the model's architecture \cite{hu2021loralowrankadaptationlarge}. LoRA approach reduces the computational overhead and memory requirements of fine-tuning large language models. Direct Preference Optimization (DPO) stands in contrast by directly refining the parameters of the language model to enhance the probability of producing text that aligns closely with the given preference data. This approach streamlines the tuning process by eliminating the need for additional external steps, such as a separate reward model or Reinforcement Learning from Human Feedback \cite{rafailov2023direct}. An additional enhancement to LoRA is quantized low rank adaptation (QLoRA), which incorporates a quantization mechanism that compresses model parameters into a 4-bit format. This approach further decreases VRAM requirements, albeit with a trade-off of increased computational time and a minor reduction in accuracy \cite{dettmers2023qloraefficientfinetuningquantized}. Finally, Quantized Aware Training (QAT) is an extension of DPO that incorporates simulating quantization numerics during fine-tuning. This allows the model to adjust for quantization noise when updating the weights, which aims to improve memory-saving optimizations at inference time, without significantly degrading model performance, during later model quantization \cite{hasan2024optimizinglargelanguagemodels}.

The choice between LoRA and DPO for fine-tuning LLMs often depends on specific goals and constraints of the application. For scenarios where computational efficiency is crucial but flexibility isn't paramount, LoRA could be a preferred method due to its efficient parameter usage. As computing performance is the main focus of this work, it has been chosen for LLM tests.

\subsubsection{Methodology for LLMs fine-tuning}

The tuning process was carried out with TorchTune, employing the recipes designed for both single-device and distributed Llama3-8B LoRA. Subsequently, these recipes were adjusted through custom configuration files to facilitate the execution of required test variants. These configuration files encompass parameters that govern and enhance the tuning procedure. For the following parameters, the default values were used - batch size: \texttt{2}, learning rate: $3\mathrm{e}{-4}$, warm-up steps: \texttt{100}, epochs: \texttt{1}. 

The \textit{dataset} parameter specifies the type of dataset being used for training. Datasets used in tests are described in section \ref {subsec:datasets}. The \textit{precision} parameter determines the numerical precision of the model's weights and activations during training. When set to a higher value, this parameter enables more precise calculations, which can be beneficial for certain models or hardware configurations. However, it also increases the computational requirements and memory usage. The $pin\_memory$ parameter, used in some tests, optimizes memory allocation by pinning the dataset's memory to CPU, which can lead to improved performance of host-to-device memory operations. 

%In terms of distribution, TorchTune offers two primary modes: distributed and single-device training. Distributed training (used in this paper) utilizes Fully Sharded Data Parallel mechanism to split the model's parameters across multiple GPUs, enabling parallel processing and improved performance on large-scale datasets. 

Tuning process comprises of processing all the dataset rows in sequence. Particular rows are distributed between all available GPUs and collected in batch size group, which is processed simultaneously. One such pass constitutes an iteration in tuning process. The mean time required for its completion serves as an indicator of efficiency. After computations, results were gathered from log files. Some tests required launching with \textit{nsys} profiling enabled and further analysis using Nsight Monitor.

In the next part of experiments, a series of test datasets were created, one for each of dataset templates. Each comprises of 500 rows from parent datasets, the equal size means equal workload for measuring the impact of template type on tuning performance. Both LoRA and Direct Preference Optimization recipes were configured to conduct exact same conditions for each run - operating in full precision mode (\texttt{FP32}), a learning rate of $3\mathrm{e}{-4}$, dataset shuffling enabled, batch size set at 2, single training epoch, gradient accumulation steps fixed at 1, activation checkpointing disabled and memory efficient Fully Sharded Data Parallel (fsdp) wrapping activated. Finally, running times and profiling data were evaluated for these tasks.

\subsection{Investigated datasets} \label{subsec:datasets}
In order to properly understand the performance impact of tuning, it is first necessary to understand the differences between datasets used. Details of selected datasets from the \textit{Torchtune} library are presented below. Each one uses a particular template, which are used to format prompts to optimize model performance on specific tasks, e.g. answering questions, summarizing or correcting errors. Each template includes the template prompt with placeholders for the data inputs.

\textbf{\underline{alpaca-cleaned}} - this is a cleaned version of the original Alpaca Dataset released by Stanford, which was generated by a language model \textit{text-davinci-003}. This instruction data is designed for instruction-tuning for pertrained language models. Due to a generated nature of the data, which inevitably contained some errors or biases, the following issues have been identified and fixed such as: hallucinations, merged instructions, empty outputs, empty code examples, N/A outputs, wrong answers, non-sensical or unclear instructions, and extraneous escape and control characters. The dataset consists of 51760 rows with the following fields: \textit{instruction}, which commands the model what it should do, \textit{input} an optional context, which is replaceable for the particular task and finally \textit{output} containing the result expected. The tasks consist of answering questions in a concise manner, using built-in knowledge and reasoning, including classification, instruction following, and writing. The template used for tuning is InstructTemplate.

\textbf{\underline{grammar}} - the dataset (C4\_200M Synthetic Dataset for Grammatical Error Correction) is a collection of 185 million sentence pairs generated from the cleaned English dataset from C4 public Common Crawl web scrape. It is the largest of tested datasets and can be used in grammatical error correction (GEC) tasks. Each dataset row includes two fields: \textit{input}, containing incorrect sentence and \textit{output}, consisting of revised phrase. An example line may look like this:
  "input": \texttt{"Bitcoin is for \$7,094 this morning, which CoinDesk says."}
  and "output": \texttt{"Bitcoin goes for \$7,094 this morning, according to CoinDesk."}

As evidenced, the tasks assigned include finding words or phrases that contain language errors and replace them with correct form. The template used for tuning is \texttt{GrammarErrorCorrectionTemplate}. The corruption edits and scripts used to synthesize this dataset is referenced from BEA 2021 paper\cite{stahlberg-kumar-2021-synthetic}. 

\textbf{\underline{samsum}} - the dataset is the smallest of tested datasets and was created to emulate conversation styles and topics commonly encountered in messaging apps. It contains about 16,369 conversations annotated with summaries Each conversation includes a unique identifier and detailed metadata such as dialogue text, summary, speaker names, and more. Conversations vary in formality, containing diverse language including slang, emoticons, and typos. The annotations are designed as concise briefs of the discussed topics from a third-person perspective. This dataset supports tasks related to summarization in conversational data across various platforms, including chatbot dialogues, SMS, IRC/chat, movie scenes, tweets, and daily communications. The creation process involved linguists drafting conversational content similar to their daily written exchanges, followed by language experts annotating these conversations for key information extraction in third-person form. This dataset is aimed at advancing research on summarization techniques applicable across multiple natural language processing (NLP) tasks and platforms, facilitating the development of AI models capable of generating coherent and contextually relevant summaries from conversational data \cite{gliwa-etal-2019-samsum}. The template used for tuning is SummarizeTemplate.

%m{6cm}|
\begin{table}[h]
\centering
\caption{Dataset characteristics}
\label{tab:dataset_summary}
{\renewcommand{\arraystretch}{1.4}
\adjustbox{max width=\linewidth}{
\begin{tabular}{|l l l m{3.5cm} m{2.5cm}|}
\hline
Dataset & Size (MB) & Length (pairs) & Tasks & Template\\ \hline
\texttt{alpaca-cleaned} & 24.1 & 51,760 & Classification, summarization, and writing & Instruct \\  \hline
\texttt{grammar} & 3710.5 & 185,000,000 & Correcting grammatical errors & Grammar Error Correction  \\  \hline
\texttt{samsum} & 10.71 & 16,369 & Conversation summarization & Summarize   \\  \hline
%SlimOrca & 307 & 363,491 & Language modeling, text generation, and augmentation & Instruct  \\ \hline
\end{tabular}
}
}
\end{table}
\vspace{10pt}

\subsection{Evaluation}
All tests in this section were performed using the same value for the following parameters: batch size:\textit{2}, learning rate: $3\mathrm{e}{-4}$, warm-up steps: \texttt{100}, epochs: \texttt{1}. Detailed information is provided as description to each relevant test.

\subsubsection{Optimization results}
\begin{figure}[h]
%źródło: arkusz - LLM/Energy, memory, iteration time results; skoroszyt - wykres
    \centering
    \includegraphics[width=0.65\linewidth]{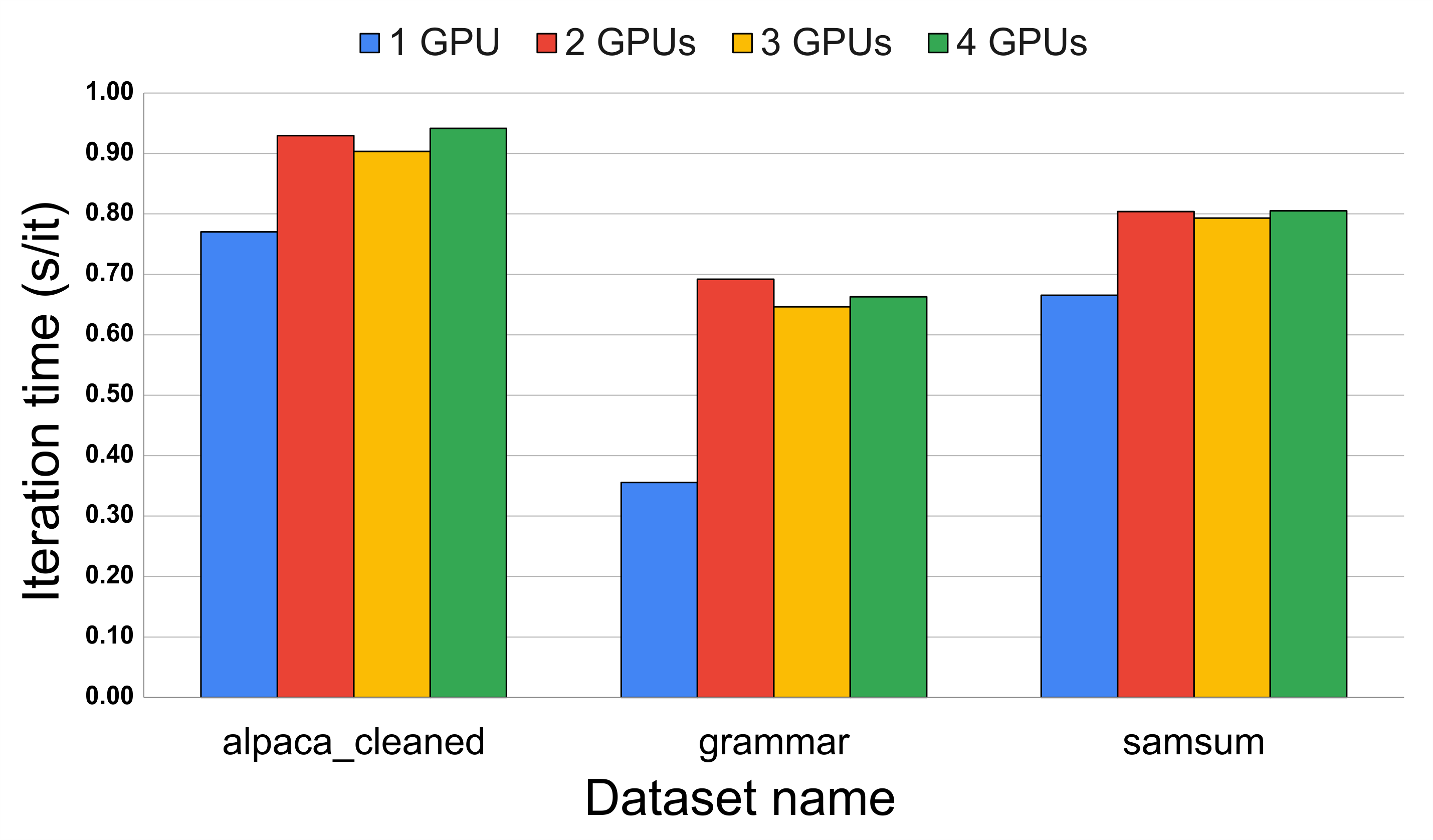}
    \caption{Dataset tuning performance scaling}
    \label{fig:llmiterationtime}
\end{figure}

The performance impact of the tuned dataset was measured by determining the average time required to complete an iteration—a group of dataset rows processed simultaneously, the size of which depended on the number of GPUs and the batch size during LoRA tuning.
Figure \ref{fig:llmiterationtime} depicts average iteration times for LLM tuning for specified datasets: \texttt{alpaca-cleaned}, \texttt{grammar} and \texttt{samsum}, compared when scaling using 1, 2, 3, and 4 GPUs. The experiments were conducted using \texttt{FP32} precision with $pin\_memory$ option disabled. 

The results indicate that there is a slight increase in average iteration time when adding more GPUs for all datasets. This can be explained by additional communication and synchronization between GPUs needed for finishing each iteration. \texttt{Grammar} and \texttt{samsum} appear to have shorter iteration time, whereas \texttt{alpaca-cleaned} has the longest. In addition adding more than 2 GPUs appears to have little effect on iteration time, what can be associated with the additional cost resulting from inter-socket communication. Of course, utilizing more GPUs reduces the number of iterations necessary to process the whole dataset, which in turn cuts total time to fine-tune.

%\subsubsection{Data operations analysis}
All runs testing \texttt{$pin\_memory$} flag were using LoRA fine-tuning on \texttt{alpaca-cleaned} dataset with \texttt{FP32} precision and \textit{batch size} set to 2.
Table \ref{tab:pinmemorytime} presents a comparison between tuning time run on 1,2,3 or 4 GPUs with and without $pin\_memory$ flag enabled. There is very little variation in execution time between cases with and without this flag present. The performance gain when tuning LLM when changing this parameter seems to be negligible due to the relatively small size of the units and the low intensity of memory transfers between the CPU and GPU.

\begin{table}[ht]
%źródło: arkusz - LLM/Optimization results
    \centering
    \adjustbox{max width=\linewidth}{
    \begin{tabular}{ccccccccc}
        \toprule
        \textbf{GPUs} & \textbf{$pin\_memory$} & \textbf{Iteration time (s/it)} & \textbf{Total time} & \textbf{Relative improvement} \\
        \midrule
        1 & FALSE & 0.77 & 5:31:52 & -       \\
        1 & TRUE  & 0.77 & 5:31:44 & 0,04\% \\
        2 & FALSE & 0.93 & 3:20:30 & -       \\
        2 & TRUE  & 0.92 & 3:17:21 & 1,57\% \\
        3 & FALSE & 0.90 & 2:09:55 & -       \\
        3 & TRUE  & 0.90 & 2:09:57 & -0,03\% \\
        4 & FALSE & 0.94 & 1:41:32 & -       \\
        4 & TRUE  & 0.91 & 1:37:58 & 3,50\% \\
    \end{tabular}}
    \vspace{5pt}
    \caption{Comparison of $pin\_memory$ option performance impact.}
    \label{tab:pinmemorytime}
\end{table}

\subsubsection{Metric analysis}
Next series of tests utilize profiler data, in order to measure operations that collectively take up the most of computing time - \texttt{cudaLaunchKernel} and \texttt{cudaStreamSynchronize} and memory operations \texttt{Host-to-Device}. The tests were performed  during LoRA fine-tuning on different datasets, using a single GPU and \texttt{FP32} precision. Because tuning process can take up considerable time, which is also variable between datasets, in order to make a just comparison, profiling results were collected after first 12 minutes (720s) of processing in all cases. Additionally, \texttt{nccldev\_AllGather} metric is not included, as it is only executed during distributed tuning (using more than 1 GPU). 

\iffalse
\begin{table}[ht]
%źródło: arkusz - LLM/nsys; skoroszyty - proxima alpaca, proxima grammar, proxima slimorca, proxima samsum
    \centering
    \adjustbox{max width=\linewidth}{
    \begin{tabular}{lcccc}
        \toprule
        \textbf{Dataset} & \textbf{API call} & \textbf{Time(\%)} & \textbf{Total Time (s)} & \textbf{Iteration time (s/it)}\\
        \midrule
        alpaca-cleaned & cudaLaunchKernel & 19.8\% & 227.973 & 0.77 \\
        grammar & cudaLaunchKernel  & 13.6\% & 134.447 & 0.36 \\
        %slimorca & cudaLaunchKernel & 20.9\%  & 246.633 & 1.23 \\
        samsum & cudaLaunchKernel  & 18.9\%  & 216.052 & 0.67 \\
        alpaca-cleaned & cudaStreamSynchronize & 23.30\% & 268.144 & 0.77 \\
        grammar & cudaStreamSynchronize  & 25.30\% & 250.179 & 0.36 \\
        %slimorca & cudaStreamSynchronize & 21.0\% & 248.012 & 1.23 \\
        samsum & cudaStreamSynchronize  & 23.8\% & 271.527 & 0.67 \\
        alpaca-cleaned & Host-to-Device & 0.10\% & 1.054 & 0.77 \\
        grammar & Host-to-Device  & 0.10\% & 1.054 & 0.36 \\
        %slimorca & Host-to-Device & 0.10\% & 1.049 & 1.23 \\
        samsum & Host-to-Device  & 0.10\% & 1.057 & 0.67 \\
    \end{tabular}}
    \vspace{5pt}
    \caption{Comparing \texttt{cudaLaunchKernel} and \texttt{cudaStreamSynchronize} time for different datasets.}
    \label{tab:datasetapi}
\end{table}
\fi

\begin{figure}[h]
%źródło: arkusz - LLM/nsys; skoroszyt - wykres dataset
    %\centering    
    \includegraphics[width=0.95\linewidth]{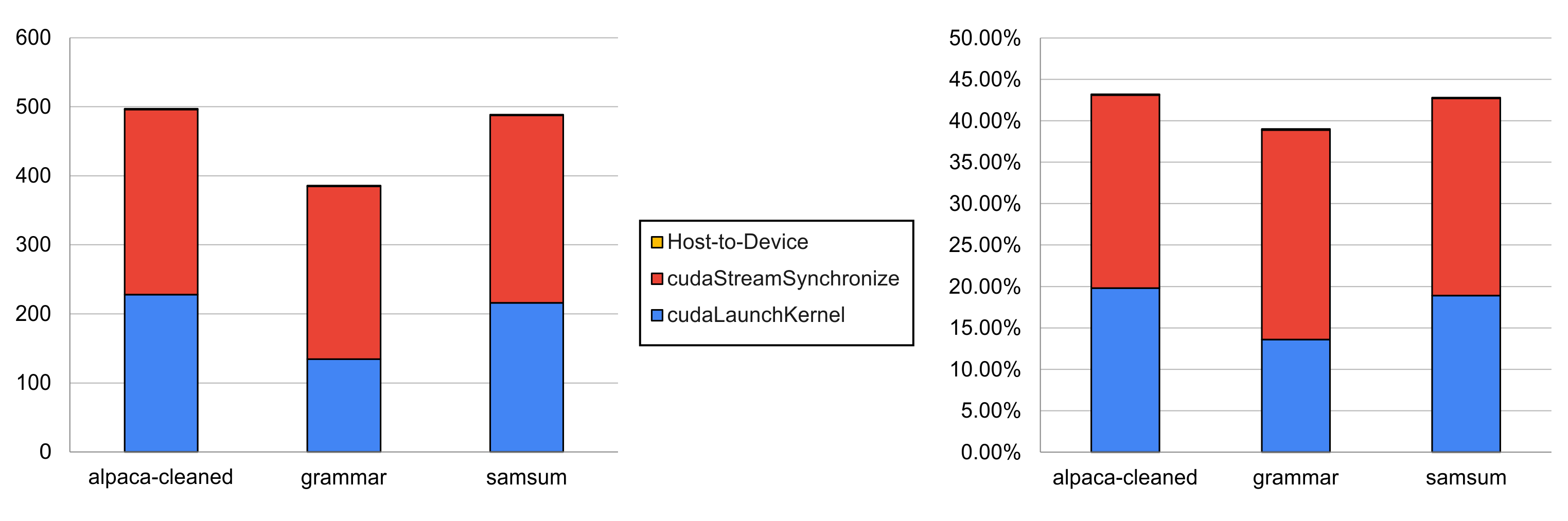}
    \caption{Cuda API calls utilization for tested datasets in seconds (left) and in percentage (right)}
    \label{fig:llmdatasetapi}
\end{figure}

As can be noted in figure \ref{fig:llmdatasetapi} the general trend appears to be that significant amount of processing time is spent on launching new kernels and synchronizing numerous threads. This can be interpreted by the multitude of iterations, that need to be sequentially executed in order to complete the tuning process. Additionally, total time spent on executing \texttt{cudaLaunchKernel} operations seems to correlate with average iteration time of the dataset. Conversely, with \texttt{cudaStreamSynchronize} calls, there is much less variation in total time and no such correlation occurs. Therefore, it is concluded that communication operations remain relatively constant on a single GPU, regardless of dataset size or its tasks, and time spent on launching new kernels is related to the speed of iteration processing. Memory operations do not seem to have a significant impact on processing time in any case. In case of API call tests to the $pin\_memory$ flag, similarly to the case of dataset measurements, profiling results were collected after the first 12 minutes of processing.

\begin{figure}[h]
%źródło: arkusz - LLM/nsys; skoroszyt - pin wykres 
    %\centering
    \includegraphics[width=0.5\linewidth]{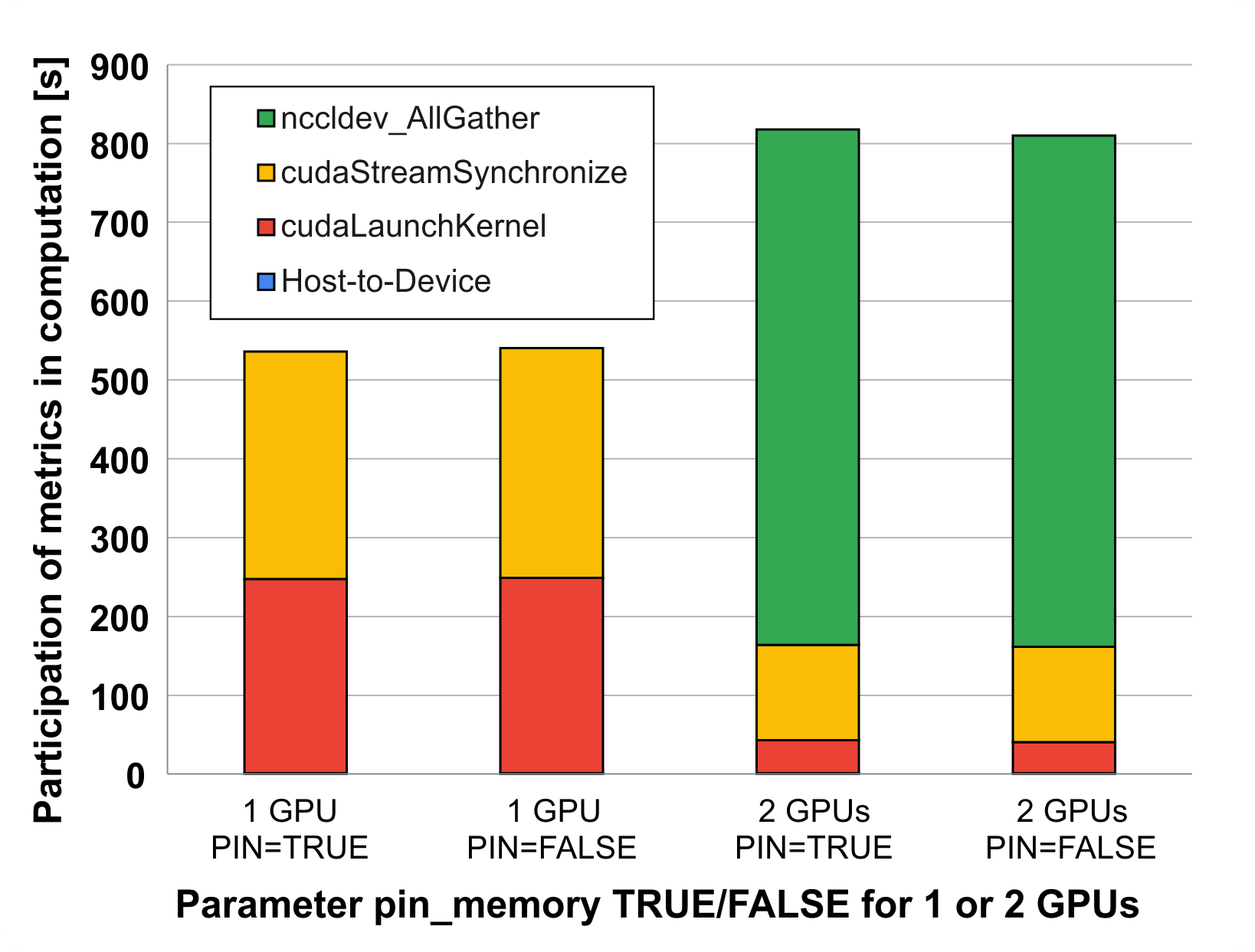}
    \includegraphics[width=0.5\linewidth]{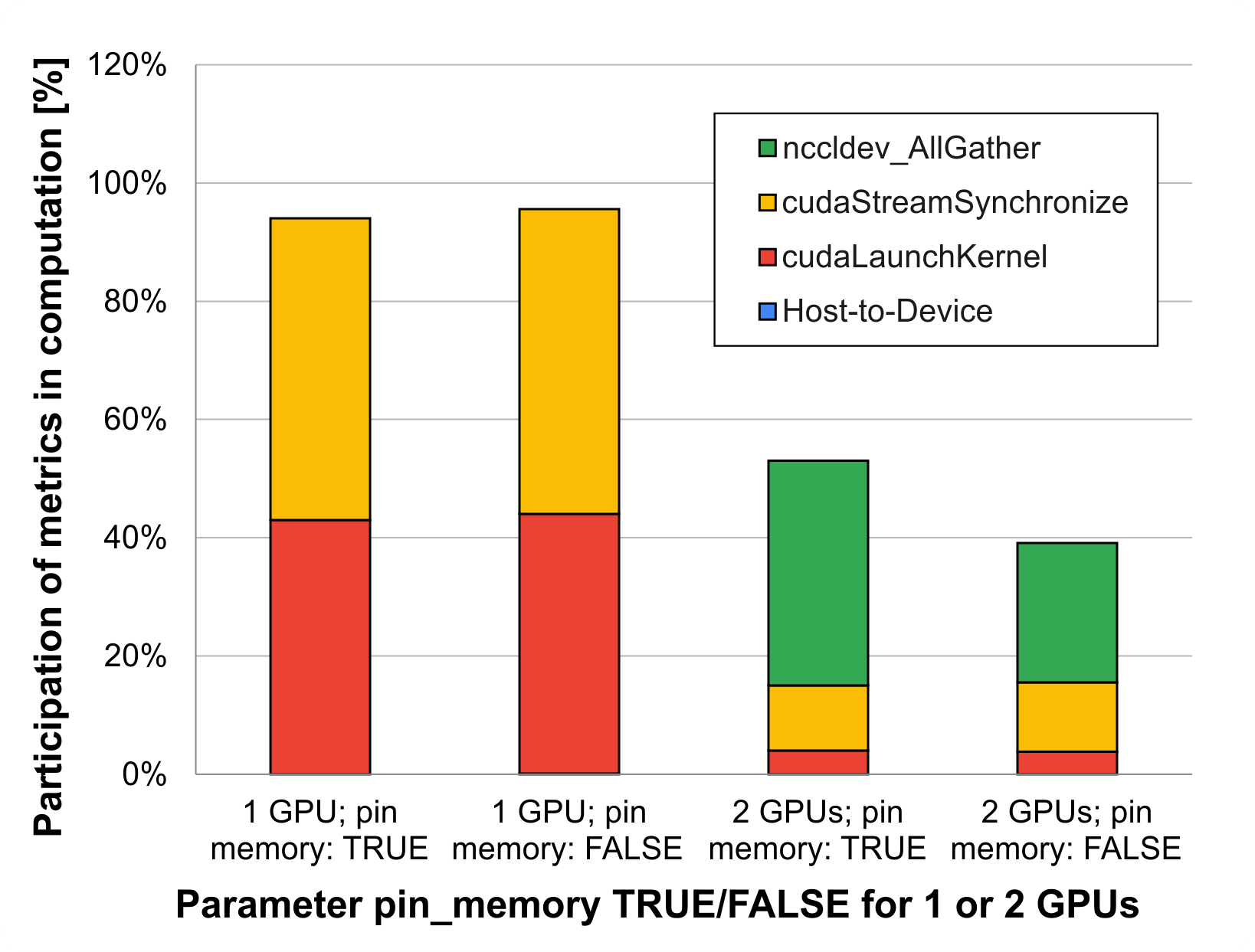}
    \caption{Cuda API calls time for $pin\_memory$ in seconds (left) and percent (right) distributed on 1 or 2 GPUs}
    \label{fig:llmpinmemoryapi}
\end{figure}

Chart displayed in figure \ref{fig:llmpinmemoryapi} presents profiling data from API calls with respect to \texttt{pin memory} option. In addition to \texttt{cudaLaunchKernel}, \texttt{Host-to-Device} memory transfers were measured, as well as \texttt{nccldev\_AllGather} in cases where tuning was distributed among multiple GPUs. The collection of profiling data from all accelerators has resulted in the possibility of the total sum exceeding 720 seconds due to concurrency. When $pin\_memory$ is enabled, there is a minor increase in the already negligible Host-to-Device memory operations. Beyond this observation, no significant performance improvements are evident from the application of this optimization technique.

In conclusion, the utilization of contemporary hardware indicates that even the most extensive datasets do not produce a sufficient number of memory operations to substantially influence the tuning process. This restricted quantity of memory operations can be attributed to the relatively small size of the dataset in comparison to the model's size, as well as the fact that training is conducted solely over a single epoch, resulting in each row of the dataset being utilized only once.

\subsubsection{Template analysis}

The following set of tests were performed on a set-length (500 rows) extracts from larger datasets, one for each template. Equal-sized subsets are used to ensure a fair comparison of workloads, which allows measuring the impact of different templates on tuning outcomes. However, the total size of respective resulting datasets is not identical, but is as follows - Instruct: 59330 tokens (\texttt{374kB}), Summary: 83544 tokens (\texttt{341kB}), Error Correction: 28683 (\texttt{135kB}). The token differences are due to row lengths.

\begin{figure}[h]
%źródło: arkusz - LLM/lora, qlora, full; skoroszyt - time  (na dole); 4gpu - LLM/qat; skoroszyt time results
    %\centering
    \includegraphics[width=0.5\linewidth]{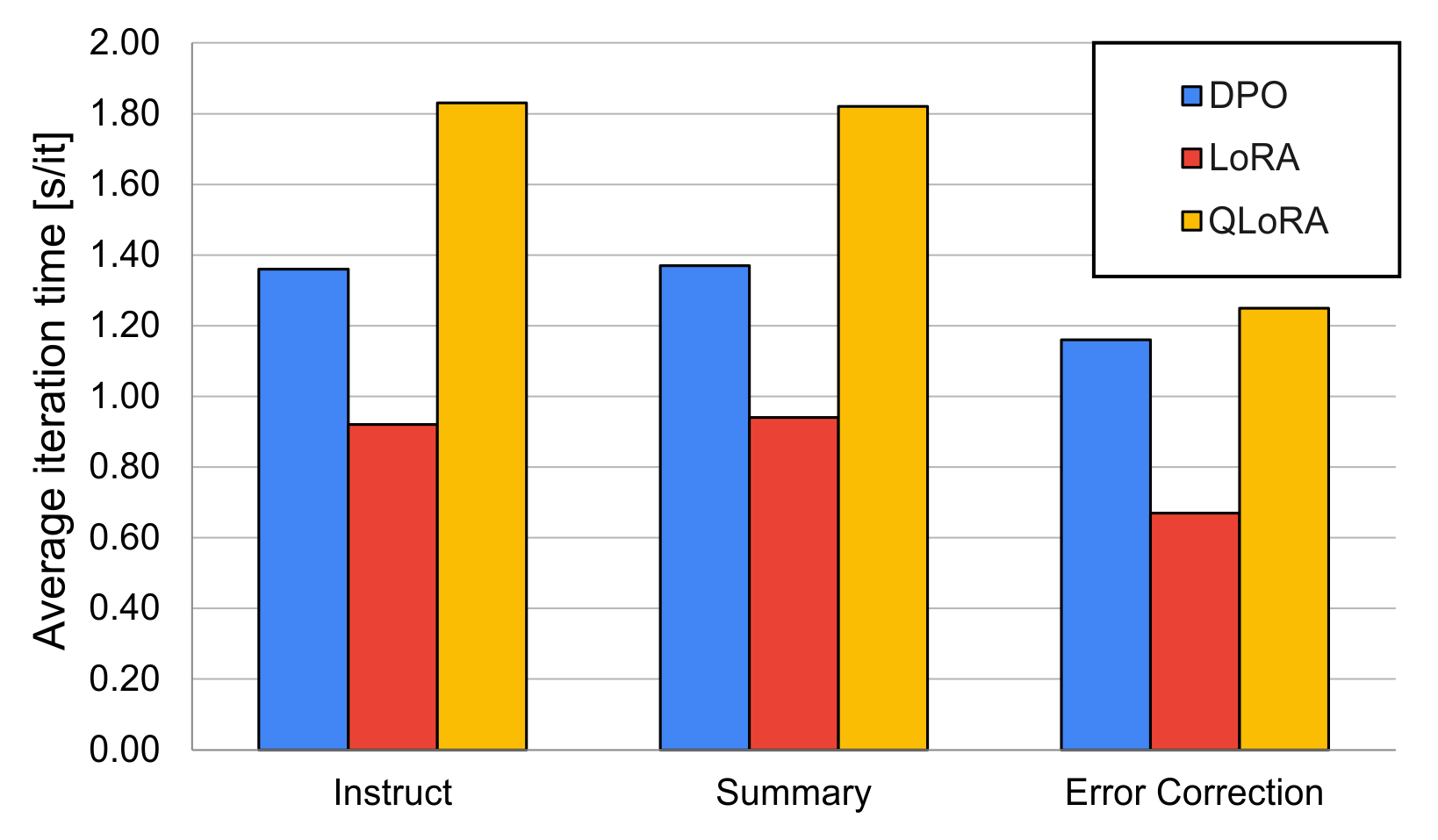}
    \includegraphics[width=0.5\linewidth]{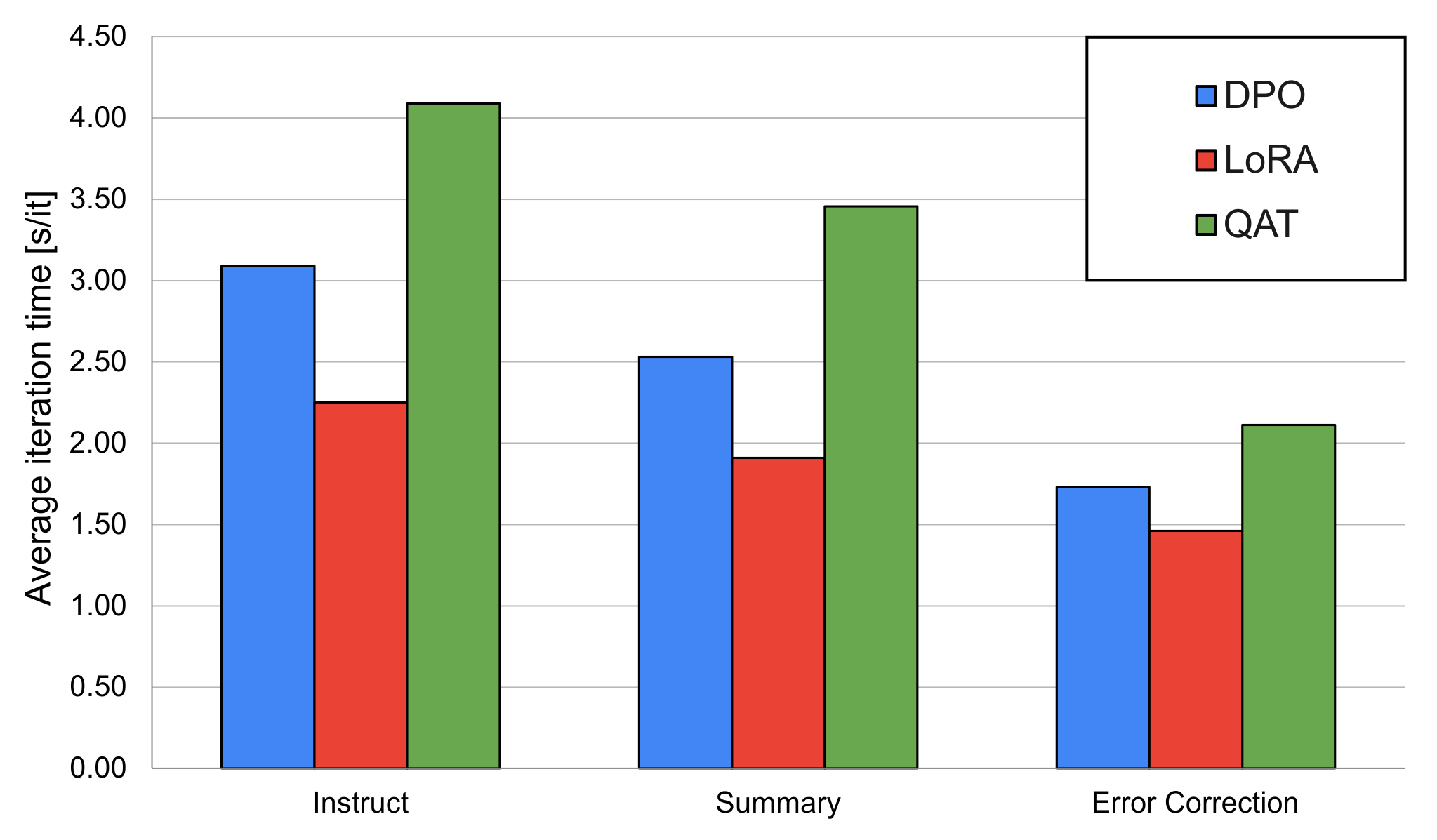}
    %\caption{Single GPU template and tuning type time comparison}
    \caption{Single GPU (left) and 4 GPUs (right) templates and tuning types time comparison}
    \label{fig:llmtemplate1gpu}
\end{figure}

In those experiments all approaches - QLoRA, LoRA, Direct Preference Optimization and Quantized Aware Training - had their recipes configured to exact same values. This means utilizing full precision mode (\texttt{FP32}), a learning rate of $3\mathrm{e}{-4}$, dataset shuffling \texttt{enabled}, batch size at \texttt{2}, single epoch of tuning, gradient accumulation steps set at \texttt{1}, activation checkpointing \texttt{disabled} and memory efficient Fully Sharded Data Parallel (fsdp) wrapping turned \texttt{on}. The rest of configuration is left at default Torchtune values. Figure \ref{fig:llmtemplate1gpu} shows average iteration time for tuning datasets with particular template on a single GPU and 4 GPUs. This way it is possible to showcase performance differences in two particular instances - single device, where it is possible to run QLoRA recipe as well as multi-node tuning with enough video memory to be able to run QAT tuning.

\begin{figure}[h]
%źródło: arkusz - LLM/lora, qlora, full; skoroszyt - time (na dole)
    %\centering
    \includegraphics[width=0.5\linewidth]{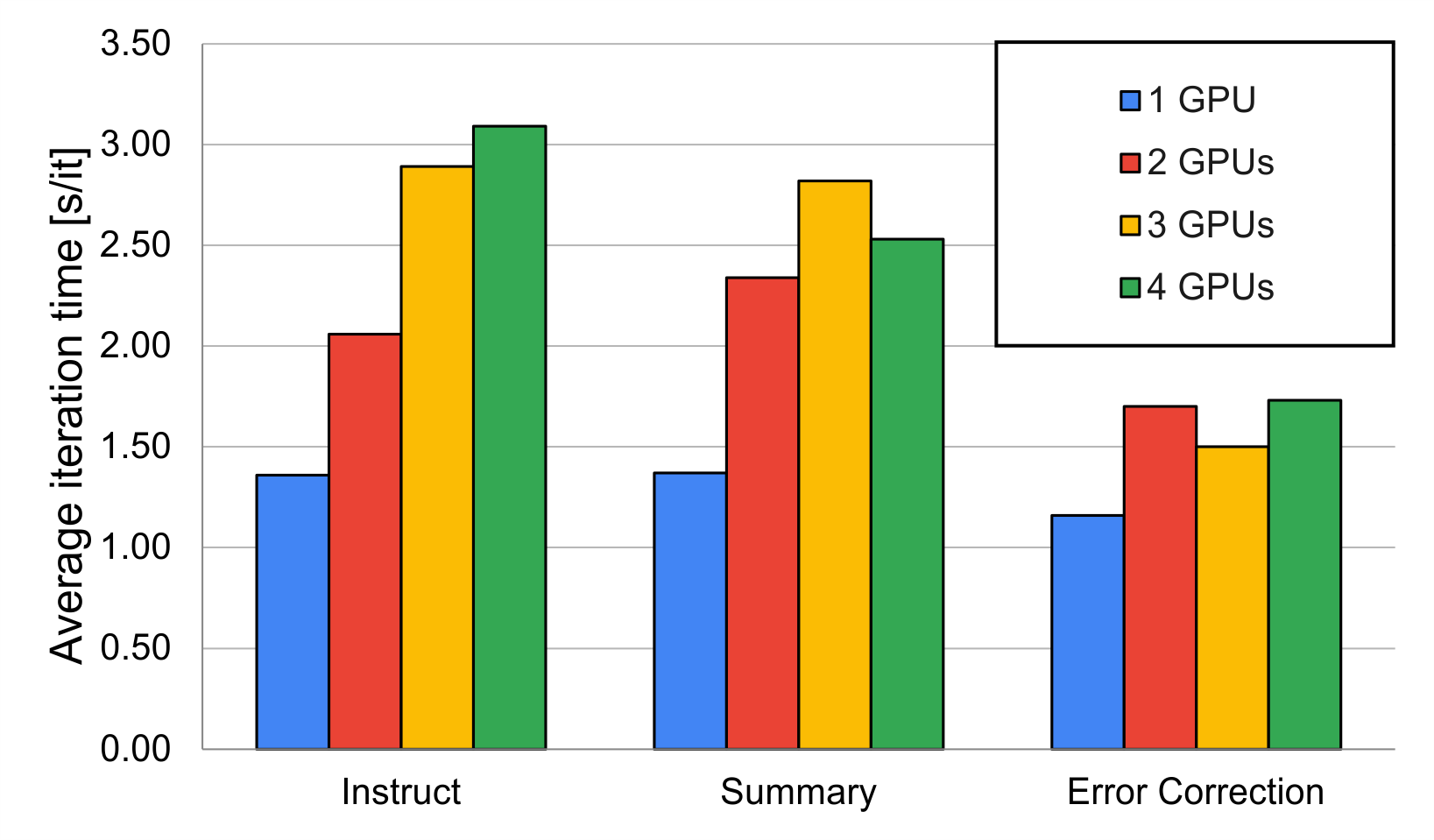}
    \includegraphics[width=0.5\linewidth]{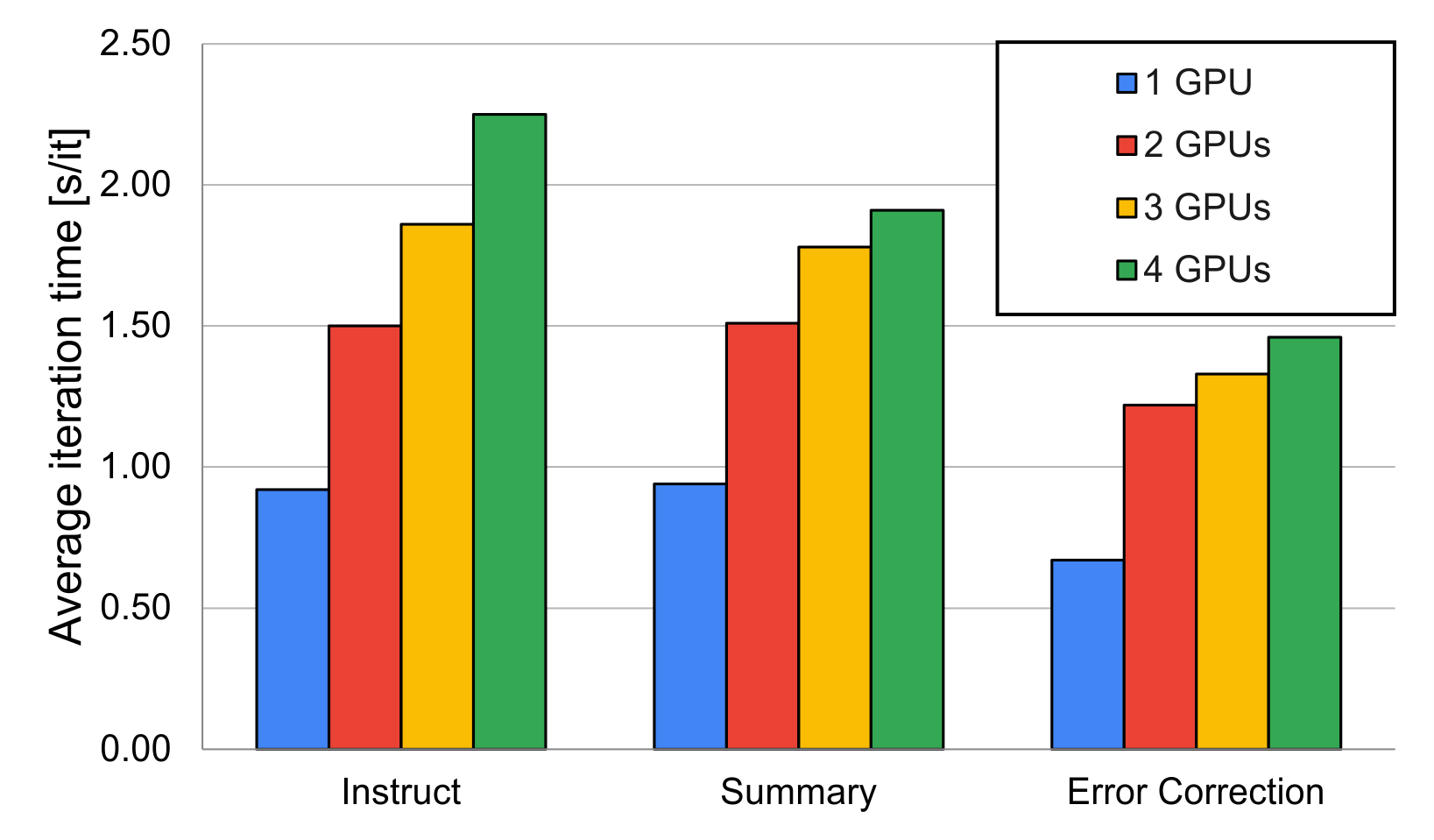}
    \caption{DPO (left) and LoRA (right) performance scaling with different dataset templates}
    \label{fig:llmtemplateiteraion}
\end{figure}

As you can see, the LoRA approach consistently takes the least time, while QLoRA and QAT are the slowest (they include additional steps during the process). Furthermore, the Error Correction template runs the fastest, while Summary and Instruct are comparable in terms of performance. This is proportional to the dataset file size. QAT is even slower than DPO. Portrayed in figure \ref{fig:llmtemplateiteraion} is the comparison of template performance gains on multiple GPUs for both DPO and LoRA approaches, as they allow running both on single-device as well as distributed modes.

The times go down the more GPUs are assigned to the task. Again, tuning using Error Correction template performs the fastest and Instruct and Summary runs are similar in running time. What is also worth of note is that LoRA runs are consistently 30-40\% faster than DPO tuning. Figure \ref{fig:llmdatasetapi} presents percentage of execution time for selected CUDA API calls (cudaLaunchKernel, cudaStreamSynchronize, Host-to-Device transfers, cudaEventSynchronize, and AllGather\_RING\_LL). The results are broken down into differing dataset templates and tuning approaches performed on 4 GPUs.

\begin{figure}[h]
%źródło: arkusz - LLM/qat; skoroszyt - profiler results
    \centering
    \includegraphics[width=0.65\linewidth]{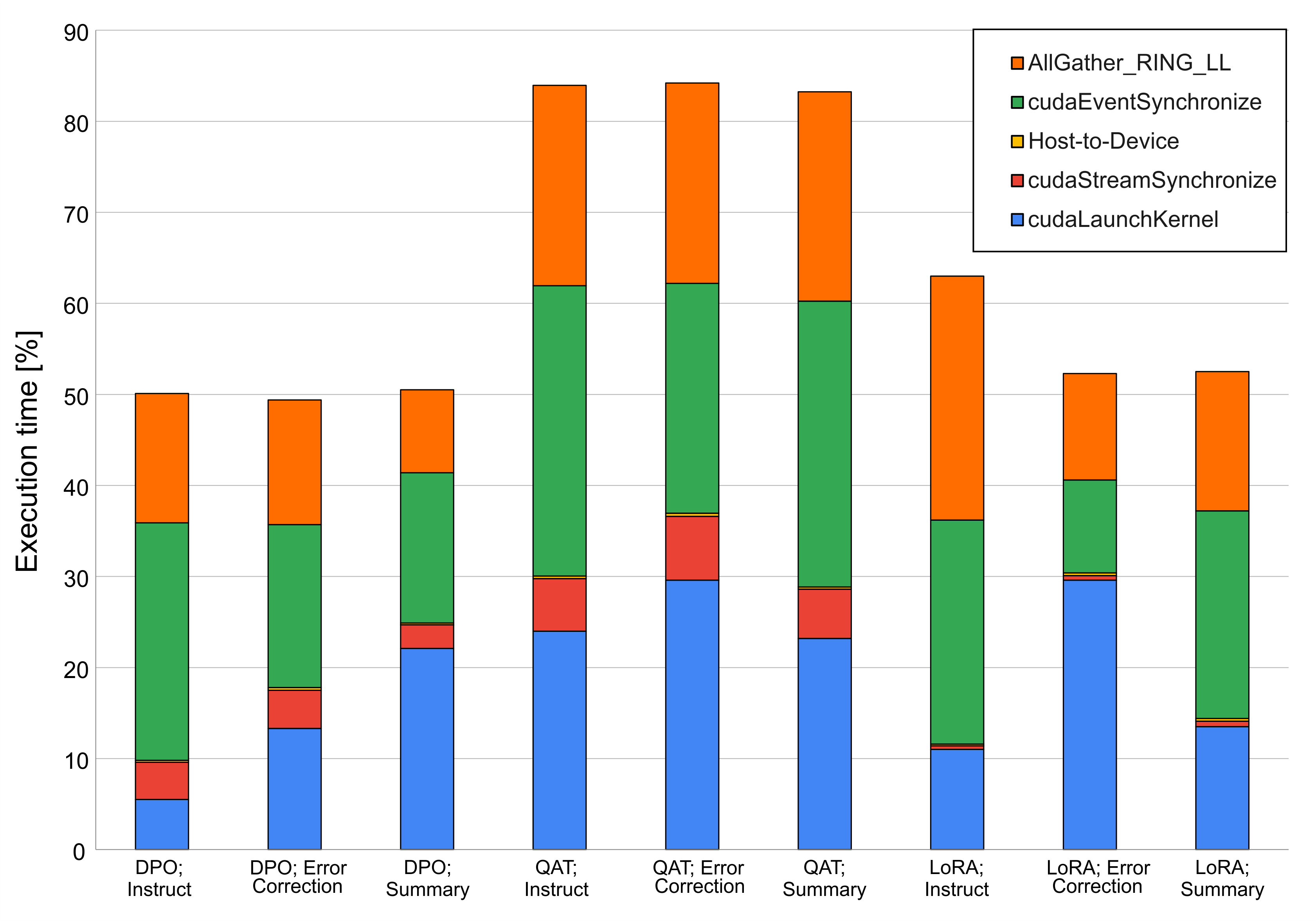}
    \caption{Cuda API calls time for different templates tuning distributed on 4 GPUs as percentage of total time }
    \label{fig:llmtemplateapi}
\end{figure}

These four API calls amount to significantly largest proportion in the case of QAT. In other instances this ratio appears to be on a similar level among different templates and tuning mechanisms, with only LoRA tuning with Instruct template having slightly higher percentage on API calls execution time in relation to total time. Across all methods, Host-to-Device transfers constitute a small fraction of the total execution time, and while task variations exist, the differences between methods are more pronounced than the differences between tasks within the same method. Among different API calls, the most prominent are \texttt{cudaEventSynchronize} and \texttt{cudaLaunchKernel}, then \texttt{AllGather\_RING\_LL}. DPO and QAT approaches make significant use of \texttt{cudaStreamSynchronize} call, which is mostly absent from LoRA profiling data. Host-to-Device memory operation's impact is negligible in every case.

\begin{figure}[h]
%źródło: arkusz - LLM/qat; skoroszyt - memory allocation
    \centering
    \includegraphics[width=0.75\linewidth]{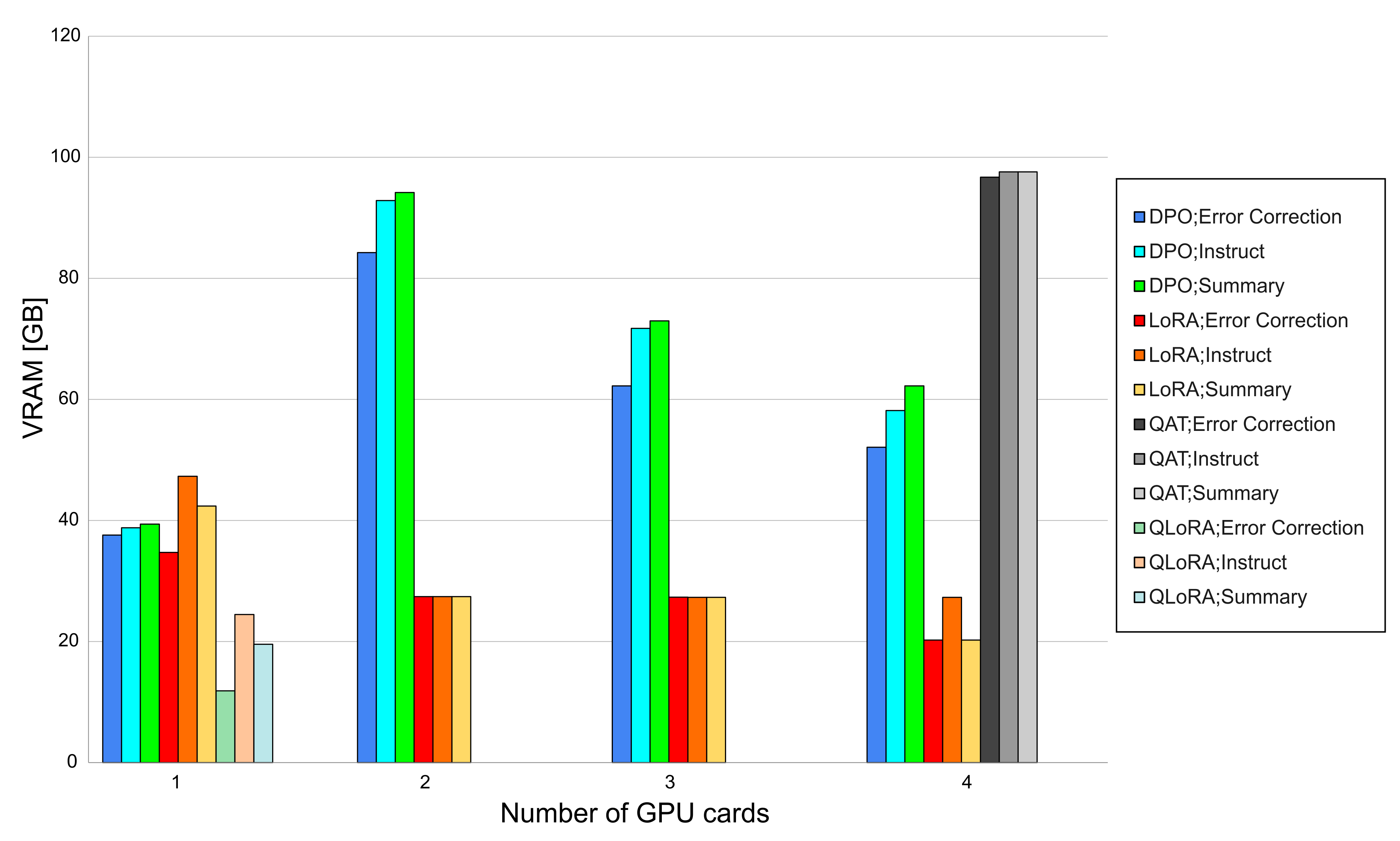}
    \caption{VRAM allocation for template tuning  on 1, 2, 3 or 4 GPUs}
    \label{fig:llmtemplatevram}
\end{figure}

Figure \ref{fig:llmtemplatevram} shows maximum VRAM allocation on a single GPU during tuning process for 1, 2, 3 or 4 GPUs for differing templates and tuning approaches. The data is collected by setting recipe options for \texttt{DiskLogger} with flag \texttt{log\_peak\_memory\_stats} and reading the last step memory allocation.

\begin{figure}[h]
%źródło: arkusz - LLM/qat; skoroszyt - memory transfers
    \centering
    \includegraphics[width=0.6\linewidth]{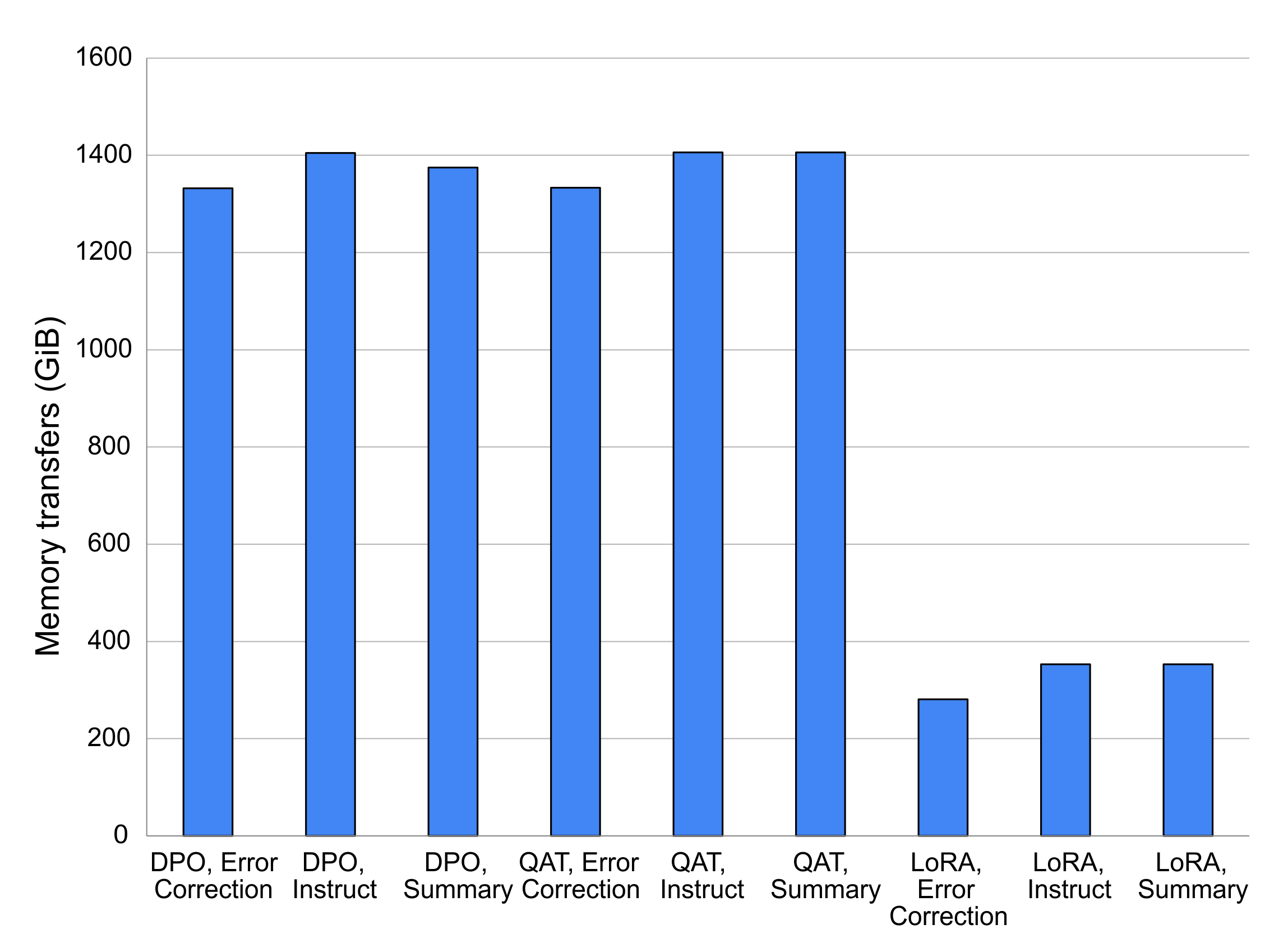}
    \caption{Total memory transfers for template tuning on 4 GPUs}
    \label{fig:llmtemplatememorytransfers}
\end{figure}

The results present several observations case by case. DPO memory requirement double when using 2 GPUs and then goes down proportionately with the number of accelerators used. LoRA does not display similar behaviour, memory consumption drops slightly when using 2 GPUs, but stays stable when adding more. QLoRA has the smallest memory requirements, although Torchtune only offers single-device tuning recipe. QAT only finished tuning on 4 GPUs as it allocates much more memory that other techniques, leading to out-of-memory errors when used with fewer than 4 GPUs. When it comes to template types, Error Correction consistently uses the least amount of memory, although often by a slight margin. Instruct template utilizes less than Summary in case of DPO tuning, but more in case of LoRA. Finally, Figure \ref{fig:llmtemplatememorytransfers} depicts the total amount of all data transfers, \texttt{Host-to-Device}, \texttt{Device-to-Device}, \texttt{Device-to-Host} for templates and tuning techniques using 4 GPUs.

What can be clearly noticed is that QAT transfers are similar to DPO which is over 4 times the amount of data flow as LoRA. Again, when it comes to template types, Error Correction creates the least traffic on data bus, whereas Instruct and Summary have similar transfer volume.  

To summarize, various templates exhibit distinct behaviours concerning performance and requirements based on the tuning mechanism employed. Testing indicates that tasks related to \texttt{Grammatical Error Correction} achieve the highest speed and necessitate the least VRAM and memory transfers, irrespective of the tuning mechanism utilized. In contrast, Instruct and Summarization tasks demonstrate comparable performance across all metrics. Regarding tuning mechanisms, the LoRA approach facilitates quicker tuning while demanding less VRAM compared to DPO. Furthermore, QLoRA sacrifices computational time to achieve reduced memory requirements. QAT, on the other hand, is the most resource-intensive method, characterized by the slowest processing times and significantly higher memory demands; these aspects must be weighed against its enhanced capability for model quantization post-tuning.

%%%%%%%%%%%%%%%%%%%%%%%%%%%%%%%%%%%%%%%%%%%%%%%%%%%%%%%%%%%%
\section{Conclusions}

In this work, the authors analyse the effectiveness of using machine learning models for image recognition and for large language models on a system with multiple GPU cards.
Image recognition tests demonstrate that adjusting numerical precision, enabling Pin Memory, and transitioning to the NHWC tensor format significantly accelerate image recognition algorithms, reducing execution time by 16–30\% when using Pin Memory and up to 210\% when switching to \texttt{FP16}. These optimizations reduce overhead associated with data transfers between the CPU and GPU, enhance computational efficiency, and minimize synchronization delays and workload distribution inefficiencies. Furthermore, the computational node architecture plays a critical role in the parallelization of such algorithms. NVIDIA DALI exhibits high resilience to NUMA architecture constraints, effectively balancing workload distribution across GPUs, even when they are connected to different NUMA sockets. In contrast, PyTorch DataLoader experiences substantial performance degradation when utilizing 3–4 GPUs, characterized by prolonged pauses before each training epoch. This suggests issues related to memory allocation and inefficiencies in data loading mechanisms within complex multiprocessor environments. An analysis of NSYS reports further reveals that modifying DataLoader to use the NHWC tensor format allows it to nearly match DALI in performance when operating with 1–2 GPUs. However, performance degradation persists when scaling to 3–4 GPUs, indicating challenges in the interaction between DDP and DataLoader in multi-GPU environments. Thus, to optimize performance in a multi-GPU setting, it is recommended employing \texttt{FP16} or \texttt{FP32}, enabling $pin\_memory$, and using DALI or an optimized NHWC-based DataLoader to improve cache utilization. Furthermore, DALI is preferred for enhanced compatibility with NUMA architectures. These strategies not only accelerate image processing but also mitigate potential bottlenecks that arise with increasing computational loads.

In the subsequent use case examined (LLM model tuning), a slight increase in the average iteration time was noted when incorporating additional GPUs (up to a maximum of two) across all datasets. Throughout the tuning process, the communication operations remain relatively stable on a single GPU, irrespective of the dataset size or the tasks involved, while the duration required to initiate new kernels is associated with the speed of iteration processing. A considerable amount of the processing time during LLM tuning is attributed to initiating new kernels and synchronizing multiple threads. Memory operations do not significantly contribute to the overall time spent on LLM tuning. The performance improvement attributed to $pin\_memory$ appears to be minimal in the context of LLM tuning.

The scope of future work includes conducting tests with higher load (larger individual specimens in respective datasets), identifying new key metrics that affect performance and searching for further universal optimization techniques such as using the DALI (NVIDIA Data Loading Library) library.

\section*{Acknowledgements}
Funded by the European Union. This work has received funding from the European High Performance Computing Joint Undertaking and Poland, Germany, Spain, Hungary, France and Greece under grant agreement number: 101093457. This publication expresses the opinions of the authors and not necessarily those of the EuroHPC JU and Associated Countries which are not responsible for any use of the information contained in this publication. The results presented in this study were prepared using the infrastructure of the Poznan Supercomputing and Networking Center.

\printendnotes

% Submissions are not required to reflect the precise reference formatting of the journal (use of italics, bold etc.), however it is important that all key elements of each reference are included.
\bibliography{references}

% \begin{biography}[example-image-1x1]{A.~One}
% Please check with the journal's author guidelines whether author biographies are required. They are usually only included for review-type articles, and typically require photos and brief biographies (up to 75 words) for each author.
% \bigskip
% \bigskip
% \end{biography}

% \graphicalabstract{example-image-1x1}{Please check the journal's author guildines for whether a graphical abstract, key points, new findings, or other items are required for display in the Table of Contents.}

\end{document}